\documentclass[11pt]{article}

\usepackage[round, sort]{natbib}
\RequirePackage[colorlinks,citecolor=blue,urlcolor=blue]{hyperref}

\usepackage{amssymb,amsmath,amsthm}
\usepackage{amsthm}
\usepackage{subfig}
\usepackage{booktabs}

\usepackage{graphicx}
\usepackage{amssymb}

\usepackage{float}

\usepackage{JASA_manu}

\newcommand{\beginsupplement}{%Counting tables and figures by S1, S2, etc.
        \setcounter{table}{0}
        \renewcommand{\thetable}{S\arabic{table}}%
        \setcounter{figure}{0}
        \renewcommand{\thefigure}{S\arabic{figure}}%
     }

\newtheorem{theorem}{Theorem}
\newtheorem{lemma}{Lemma}

\newtheorem{rem}{Remark}

{\bf}{\it}

{\bf}{\it}

\begin{document}

\title{TFisher Tests: Optimal and Adaptive Thresholding \\ for Combining $p$-Values}
\author{
Hong Zhang  \\
Department of Mathematical Sciences \\ 
Worcester Polytechnic Institute, Worcester, MA 01609  \\ 
E-mail: \texttt{hzhang@wpi.edu} \\
Tiejun Tong \\
Department of Mathematics \\
Hong Kong Baptist University, Kowloon Tong, Hong Kong \\
E-mail: \texttt{tongt@hkbu.edu.hk } \\
John Landers \\
Department of Neurology\\
University of Massachusetts Medical School, Worcester, MA 01655\\
E-mail: \texttt{John.Landers@umassmed.edu } \\
Zheyang Wu  \\
Department of Mathematical Sciences \\ 
Program of Bioinformatics and Computational Biology\\
Program of Data Science\\
Worcester Polytechnic Institute, Worcester, MA 01609  \\ 
E-mail: \texttt{zheyangwu@wpi.edu} \\
}

\maketitle

\newpage

\mbox{}
\vspace*{2in}
\begin{center}
\textbf{Author's Footnote:}
\end{center}
Hong Zhang is Doctoral Candidate, Department of Mathematical Sciences, Worcester Polytechnic Institute. Mailing address: 100 Institute Rd, Worcester, MA 01609, USA (E-mail: hzhang@wpi.edu).
Tiejun Tong is Associate Professor, Department of Mathematics, Hong Kong Baptist University. Mailing address: FSC1201, Fong Shu Chuen Building, Kowloon Tong, Hong Kong (E-mail: tongt@hkbu.edu.hk).
John Landers is Professor, Department of Neurology, University of Massachusetts Medical School. Mailing address: 55 Lake Avenue North, AS6-1053, Worcester, MA 01655, USA (E-mail: John.Landers@umassmed.edu). 
Zheyang Wu is Corresponding Author and Associate Professor, Department of Mathematical Sciences, Program of Bioinformatics and Computational Biology, and Program of Data Science, Worcester Polytechnic Institute. Mailing address: 100 Institute Rd, Worcester, MA 01609, USA (E-mail: zheyangwu@wpi.edu).
The research was supported in part by the NSF grant DMS-1309960. 

\newpage
\begin{center}
\textbf{Abstract}
\end{center}
For testing a group of hypotheses, tremendous $p$-value combination methods have been developed and widely applied since 1930's. Some methods (e.g., the minimal $p$-value) are optimal for sparse signals, and some others (e.g., Fisher's combination) are optimal for dense signals. To address a wide spectrum of signal patterns, this paper proposes a unifying family of statistics, called TFisher, with general $p$-value truncation and weighting schemes.  Analytical calculations for the $p$-value and the statistical power of TFisher under general hypotheses are given. Optimal truncation and weighting parameters are studied based on Bahadur Efficiency (BE) and the proposed Asymptotic Power Efficiency (APE), which is superior to BE for studying the signal detection problem. A soft-thresholding scheme is shown to be optimal for signal detection in a large space of signal patterns. When prior information of signal pattern is unavailable, an omnibus test, oTFisher, can adapt to the given data. Simulations evidenced the accuracy of calculations and validated the theoretical properties. The TFisher tests were applied to analyzing a whole exome sequencing data of amyotrophic lateral sclerosis. Relevant tests and calculations have been implemented into an R package {\it TFisher} and published on the CRAN. 

\vspace*{.3in}


\newpage

\section{Introduction}

The $p$-value combination approach is an important statistical strategy for information-aggregated decision making. It is foundational to a lot of applications such as meta-analysis, data integration, signal detection, etc. In this approach a group of input $p$-values $P_i, i=1, ..., n,$ are combined to form a single statistic for testing the property of the whole group. For example, in meta-analysis each $p$-value corresponds to the significance level of one study, and a group of similar studies and their $p$-values are combined to test a common scientific hypothesis. In the scenario of signal detection, each $p$-value is for one feature factor, and the $p$-values of a group of factors are combined to determine whether some of those factors are associated with a given outcome. In either scenario, regardless of the original data variation, $p$-values provide a commonly scaled statistical evidence of various sources (i.e., studies or a factors), therefore the $p$-value combination approach can be considered as combining information from different sources to make a reliable conclusion. Indeed, $p$-value combination can provide extra power than non-combination methods. In signal detection for example, weak signals could be detectable as a group but not recoverable as individuals \citep{Donoho2004, jin2014rare}. 

The question is how we should combine a given group of $p$-values. One of the earliest methods is Fisher's combination statistic proposed in 1930's \citep{Fisher1932}, which is simply the product of all $p$-values, or equivalently its monotonic log transformation: 
\begin{equation}
T = \prod_{i=1}^{n}P_{i}  \quad  \Leftrightarrow \quad  W = -2\log(T) = -2\sum_{i=1}^{n}\log (P_{i}). 
\end{equation}
Fisher's combination enjoys asymptotic optimality over any possible ways of combining $p$-values when all $p$-values represent ``signals", e.g., all studies are positive or all features are associated \citep{littell1971asymptotic,littell1973asymptotic}. In this sense, the log-transformation of Fisher's combination is superior to other transformation functions, e.g., the inverse Gaussian Z-transformation \citep{Stouffer1949, Whitlock2005}. However, in real applications, it is often the case that only part of the $p$-values are related to signals.  One example is in the meta-analysis of differential gene expression, where the positive outcomes could happen in one or some of the studies only \citep{song2014hypothesis}. Another example is in detecting genetic associations for a group of genetic markers, where some of these markers are associated but some others are not \citep{Hoh2001, su2016adaptive}. In fact, it has been shown that when true signals are in a very small proportion, e.g., at the level of $n^{-\alpha}$ with $\alpha \in (3/4, 1)$, an optimal choice is to simply use the minimal $p$-value as the statistic \citep{Tippert1931}. However, the minimal $p$-value may no longer be optimal for denser weak signals, e.g., under $\alpha \in (1/2, 3/4)$ \citep{Donoho2004, Wu2014detection}. Thus, between the two ends of the classic methods -- the optimality of Fisher's combination for very dense signals and the optimality of minimal $p$-value method for very sparse signals -- a straightforward idea is to combine a subgroup of smaller $p$-values that more likely represent true signals. Following this idea, styles of truncation methods were proposed. For example, the truncated product method (TPM) statistic is defined as \citep{zaykin2002truncated, zaykin2007combining}: 
\begin{equation}
T=\prod_{i=1}^{n}P_{i}^{I(P_{i}\leq \tau )} \quad \Leftrightarrow \quad W = -2\log(T) =\sum_{i=1}^{n} -2 \log (P_{i}) I(P_{i}\leq \tau ),
\label{equ.TPM}
\end{equation}%
where $I(\cdot )$ is the indicator function and $\tau$ is the threshold of truncation. A variation of TPM is called the rank truncation product (RTP) method, in which $\tau$ is set as the $k$th smallest $p$-value for a given $k$ \citep{dudbridge2003rank,kuo2011novel}. 

Truncation-based methods have been widely applied in various practical studies and shown desirable performance. For example, many papers have been published in the genome-wide association studies (\cite{dudbridge2003rank, Yu2009, li2011adaptively, biernacka2011use, hongying2014modified}, and others). However, there is a lack of theoretical study on the best choice of $\tau$. Two ad hoc intuitions were considered. One is a ``natural" choice of $\tau=0.05$, the value of a typical significance level in single hypothesis test \citep{zaykin2002truncated}. The other intuition is to take $\tau$ as the true proportion of signals. In Sections \ref{Sect_Asymp_Optimality} and \ref{Sect_Simu} of this paper, however, we will show that in general neither of the two intuitions gives the best choice of $\tau$.

Moreover, even if we can get the best $\tau$ for TPM, would it be an optimal statistic? The answer is still no. In fact, besides truncation, the statistical power could be improved through properly weighting the $p$-values. In this paper, we propose a general weighting and truncation framework through a family of statistics called TFisher. We provide accurate analytical calculations for both $p$-value and statistical power of TFisher under general hypotheses. 
For the signal detection problem, theoretical optimality of the truncation and weighting schemes are systematically studied based on Bahadur Efficiency (BE), as well as a more sophisticated measure Asymptotic Power Efficiency (APE) proposed here. The results show that in a large parameter space, TPM and RTP are not optimal; the optimal method is by coordinating weighting and truncation in a soft-thresholding manner. This result provides an interesting connection to a rich literature of shrinkage and penalty methods in the context of de-noising and model selection \citep{abramovich2006adapting, donoho1995noising}.

When prior information of signal patterns is unavailable, an omnibus test, called oTFisher, is proposed to obtain a data-adaptive weighting and truncation scheme. In general, omnibus test does not guarantee the highest power for all signal patterns, but it often provides a robust solution that performs reasonably well in most scenarios. In literature, omnibus test mostly depends on computationally intensive simulations or permutations \citep{Yu2009, li2011adaptively, lee2012optimal, lin2015test}. In order to reduce the computation and improve the stability and accuracy, we provide an analytical calculation for determining the statistical significance of oTFisher. 

The remainder of the paper is organized as follows. Problem formulation is given in Section \ref{Sect_Formulation}, where the definitions of TFisher and the settings of hypotheses are clarified. For the whole TFisher family under finite $n$, we provide analytical calculations for their $p$-values in Section \ref{Sect_P_Calcu} and their statistical power in Section \ref{Sect_Calcu_Distn_Alter}. Theoretical studies of optimality based on BE and APE are given in Section \ref{Sect_Asymp_Optimality}. With extensive simulations, Section \ref{Sect_Simu} demonstrates that our analytical calculations are accurate and that our theoretical studies reflect the reality well. Section \ref{Sect_ExomeSeq} shows an application of TFisher tests to analyzing a whole exome sequencing data for finding putative disease genes of amyotrophic lateral sclerosis. Concluding remarks are given in Section \ref{Sect_Discu}. Detailed proofs of lemmas and theorems and the supplementary figures are given in Supplementary Materials.

\section{TFisher Tests and Hypotheses}\label{Sect_Formulation}

\subsection{TFisher}

With the \emph{input $p$-values} $P_i, i=1, ..., n$, the TFisher family extends Fisher's $p$-value combination to a general weighting and truncation scheme. The general formula of TFisher statistics can be equivalently written as 
\begin{equation}
T=\prod_{i=1}^{n}\left( \frac{P_{i}}{\tau_{2i} }\right) ^{I(P_{i}\leq \tau_1 )} \quad \Leftrightarrow \quad W=-2\log T=\sum_{i=1}^{n} \left( -2 \log (P_{i}) + 2\log (\tau_{2i}) \right)I(P_{i} \leq \tau_1 ) , 
\label{equ.W}
\end{equation}
where $\tau_1$ is the truncation parameter that excludes too big $p$-values and $\tau_{2i}$ are the weighting parameters for $p$-values. This statistic family unifies a broad range of $p$-value combination methods. When $\tau_1 = \tau_{2i} =1$, the statistic is the traditional Fisher's combination statistic. When  $\tau_1 \in (0, 1)$ and $\tau_{2i} =1$, it becomes the truncated product method (TPM) \citep{zaykin2002truncated}. When $\tau_1 = P_{(k)}$ and $\tau_{2i} =1$ for a given $k$, where $P_{(1)} \leq ... \leq P_{(n)}$ are the ordered input $p$-values, it becomes the rank truncation product method (RTP) \citep{dudbridge2003rank}. When $\tau_1 = 1$ and $\tau_{2i}=P_i^{1-\lambda_i}$, it leads to the power-weighted $p$-value combination statistic $T = \prod_{i=1}^{n}P_{i}^{\lambda _{i}}$ \citep{Good1955, li2011adaptively}. For the simplicity of theoretical studies, in what follows we restrict to constant parameters $\tau_1$ and $\tau_{2i} = \tau_2$. Such two-parameter definition corresponds to the dichotomous mixture model in the classic signal detection setting, such as those specified in (\ref{equ.GaussianHypo}) and (\ref{equ.mixtureModel}). %, where the alternative is a mixture model of dichotomous statuses. 

\begin{figure}[h] \centering%
\begin{tabular}{l}
\includegraphics[width=3in]{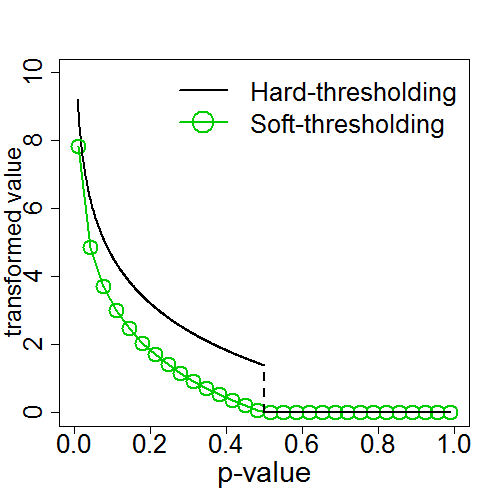}
\end{tabular}%
\caption{Comparison between the hard-thresholding curve $-2\log (P_{i}) I(P_{i} \leq \tau)$ (black) and the soft-thresholding curve $2 \left( -\log (P_{i})+\log (\tau )\right) _{+}$ (green dot). $\tau=0.5$. }
\label{fig. softvshard}
\end{figure}

The weighting and truncation scheme is also related to thresholding methods in a rich literature of shrinkage estimation, de-noising and model selection \citep{donoho1995noising, abramovich2006adapting}. In particular, when $\tau_1 = \tau$ and $\tau_2=1$, TFisher corresponds to the hard-thresholding (i.e., TPM): 
\begin{equation}
W_h = \sum_{i=1}^{n}\left( -2\log (P_{i})\right)I(P_{i} \leq \tau_1 ).
\end{equation}
When $\tau_1=\tau_2=\tau$,  TFisher is a soft-thresholding method:
\begin{equation}
W_s = \sum_{i=1}^{n}\left( -2\log (P_{i})+2\log (\tau )\right) _{+},
\label{equ.TPMsoft}
\end{equation}%
where $\left( x\right) _{+}=\max \left\{ x,0\right\} $. The soft-thresholding could have three benefits over hard-thresholding here. First, a value $\tau_2 \in (0, 1)$ downscales the significance of original $p$-values, which could reduce the type I error rate in the related context of multiple hypotheses testing. Secondly, even though $E_{H_1}(W_s) - E_{H_0}(W_s) < E_{H_1}(W_h) - E_{H_0}(W_h)$, $W_s$ has a much smaller variance, which could make itself more powerful than $W_h$. Thirdly, $W_s$ has a better weighting scheme for small $p$-values. To see this point, 
Figure \ref{fig. softvshard} illustrates that the hard-thresholding scheme, represented by the curve $-2\log (P_{i}) I(P_{i} \leq \tau)$, is discontinuous at the cutoff $\tau$. In contrast, the soft-thresholding scheme $2 \left( -\log (P_{i})+\log (\tau )\right) _{+}$ is pushed down to be a smoothed curve. The more steeply dropping curve of the soft-thresholding gives relatively heavier weights to smaller $p$-values that are more likely associated with true signals.  
In Section \ref{Sect_Asymp_Optimality}, we will provide theoretical result for functional relationships between signal patterns and optimal $\tau_1$ and $\tau_2$. The soft-thresholding is to be shown mostly optimal, which is consistent with the conclusion in shrinkage analysis \citep{donoho1995noising}.

When there is no prior information on the signal pattern, the optimal $\tau_1$ and $\tau_2$ are difficult to determine. However, we can apply an omnibus test, called oTFisher, which adapts the choice of these parameters to the given data. oTFisher does not guarantee the highest power, but it often provides a robust test that performs reasonably well over most signal patterns. In general, oTFisher adaptively chooses $\tau_1$ and $\tau_2$ that give the smallest $p$-value over the space of $(0, 1] \times (0, +\infty)$:
\begin{equation*}
W_o = \min_{\tau_1, \tau_2} G_{\tau_1, \tau_2}(W(\tau_1, \tau_2)),
\end{equation*}
where $G_{\tau_1, \tau_2}$ is the survival function of $W(\tau_1, \tau_2)$ defined in (\ref{equ.W}) under the null hypothesis. For practical computation, we study a discrete domain over $(\tau_{1j}, \tau_{2j})$ for $j=1, ..., m$:
\begin{equation}
W_o = \min_j G_j(W_j).
\label{equ.W_o}
\end{equation}
As we will show in theory and in simulations, a grid of $\tau_{1j} = \tau_{2j} \in (0, 1)$ over small, mediate and large values in $(0, 1)$ could perform sufficiently well in most cases. 

\subsection{Hypotheses}

To answer the key question of how $p$-values should be combined, we keep in mind that the performance, in particular the statistical power, of different methods depends on the setting of the null and alternative hypotheses. A general setting for the group testing problem is given in the following. For independent and identically distributed (i.i.d.) \emph{input statistics} $X_1, ..., X_n$, we aim at testing the null and alternative hypotheses:
\begin{equation}
 H_0: X_i \sim F_0 \text{ for all $i$} \quad \text{ vs. } \quad H_1: X_i \sim F_1 \text{ for all } i, 
  \label{equ.GeneralHypo}
\end{equation}
where $F_j$, $j=0,1$, denote arbitrary continuous cumulative distribution functions (CDFs). Based on the given $H_0$, the corresponding input $p$-values are
\begin{equation}
P_i = \bar{F}_0(X_i), 
\label{equ.P}
\end{equation}
where $\bar{F}_0 = 1- F_0$ denotes the survival function of the null distribution. Note that the one-sided $p$-value definition in (\ref{equ.P}) actually covers the two-sided tests too. This is because $F_0$ is arbitrary, and the statistics can simply be replaced by $X'_i = X^2_i \sim F'_{0}$ whenever the signs of input statistics have meaningful directionality (e.g., protective and deleterious effects of mutations in genetic association studies). Also note that the i.i.d. assumption in (\ref{equ.GeneralHypo}) is for the convenience of power calculation. If $p$-value calculation of TFisher is the only concern in a data analysis, the null hypothesis can be generalized to 
\begin{equation}
	H_0: \text{Independent } T_i \sim  F_{0i} \text{, or equivalently, }  H_0: P_i \overset{{\rm i.i.d.}}{\sim} \text{Uniform}[0,1], \quad i = 1, ..., n.
\label{equ.generalH0}
\end{equation}
That is, the TFisher tests can be applied into meta-analysis or integrative analysis of heterogenous data, where input test statistics could potentially follow different distributions.  

A particularly interesting scenario is the \emph{signal detection} problem, where the target is to test the existence of ``signals" in a group of statistics. Usually the test statistics are, or can be approximated by, the Gaussian distribution. Thus the problem is to test the null hypothesis of all ``noises" versus the alternative hypothesis that a proportion of signals exist: 
\begin{equation}
 H_0: X_i \overset{{\rm i.i.d.}}{\sim} N(0, \sigma^2) \quad \text{ vs. } \quad H_1: X_i \overset{{\rm i.i.d.}}{\sim} \epsilon N(\mu, \sigma^2) + (1-\epsilon)N(0, \sigma^2), \quad i = 1, ..., n.
  \label{equ.GaussianHypo}
\end{equation}
Here the zero mean indicates the noise, the none-zero mean $\mu$ represents the signal strength, and $\epsilon \in (0, 1]$ represents the proportion of the signals. The signal patterns are characterized by the parameter space of $(\epsilon,\mu)$. For simplicity, we assume the variance $\sigma^2$ is known or can be accurately estimated, which is equivalent to assuming $\sigma=1$ without loss of generality (otherwise, data can be rescaled by $\sigma$).

%%%========================================

\section{TFisher Distribution Under $H_0$}\label{Sect_P_Calcu}

In this section we provide the calculation for the exact null distribution of TFisher in (\ref{equ.W}) when $\tau_1$ and $\tau_2$ are given. Based on that, an asymptotic approximation for the null distribution of oTFisher in (\ref{equ.W_o}) is also provided. Thus the $p$-values of TFisher and oTFisher can be quickly and accurately calculated in practical applications. 

\subsection{Exact Distribution at Given $\tau_1$ and $\tau_2$}

Consider the general null hypothesis in (\ref{equ.generalH0}). Let $U_i \overset{{\rm i.i.d.}}{\sim} \text{Uniform}[0, 1], i = 1, ..., n,$ and $N$ be the number of $U_i$ less than or equal to $\tau_1$. The TFisher statistic in (\ref{equ.W}) can be written as
\begin{equation*}
W(\tau_1, \tau_2) = \sum_{i=1}^N-2\log\left(\frac{\tau_1}{\tau_2}U_i\right).
\end{equation*}
For a fixed positive integer $k\geq 1$, it is easy to check that 
\begin{equation*}
P\left(\sum_{i=1}^k-2\log\left(\frac{\tau_1}{\tau_2}U_i\right)\geq w\right)=\bar{F}_{\chi^2_{2k}}\left(w+2k\log\left(\frac{\tau_1}{\tau_2}\right)\right),
\end{equation*} 
 where $\bar{F}_{\chi^2_{2k}}(x)$ is the survival function of a chi-squared distribution with degrees of freedom $2k$. Since $N\sim \text{Binomial}(n, \tau_1)$,
 $W$ can be viewed as a compound of this shifted chi-squared distribution and the binomial distribution:
\begin{equation*}
P(W\geq w)=(1-\tau_1)^{n}I_{\{w\leq 0\}}+\sum_{k=1}^n\binom{n}{k}\tau_1^k(1-\tau_1)^{n-k}\bar{F}_{\chi^2_{2k}}\left(w+2k\log\left(\frac{\tau_1}{\tau_2}\right)\right).
\end{equation*} 
We can further simplify the above formula by noting the relationship between $\bar{F}_{\chi^2_{2k}}(x)$ and the upper incomplete gamma function $\Gamma(s, x)$:
\begin{equation*}
\bar{F}_{\chi^2_{2k}}(x) = \int_{x}^{+\infty}\frac{u^{k-1}e^{-u/2}}{2^k(k-1)!}du=\int_{x/2}^{+\infty}\frac{y^{k-1}e^{-y}}{(k-1)!}dy=\frac{\Gamma(k,x/2)}{(k-1)!}=e^{-x/2}\sum_{j=0}^{k-1}\frac{(x/2)^j}{j!}.
\end{equation*} 

\begin{figure}[h] \centering%
\begin{tabular}{l}
\includegraphics[width=3in]{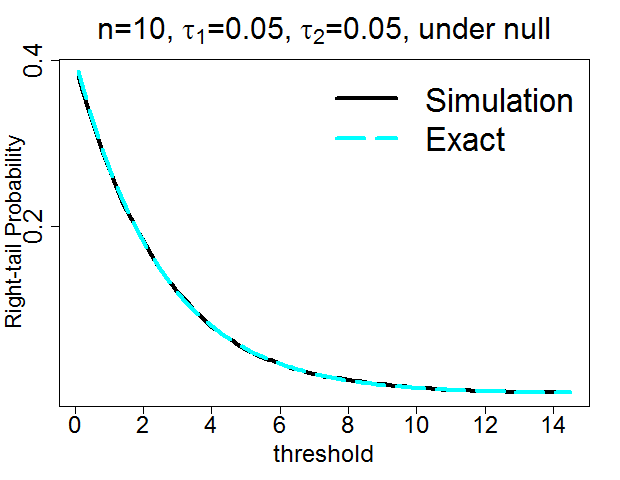}
\includegraphics[width=3in]{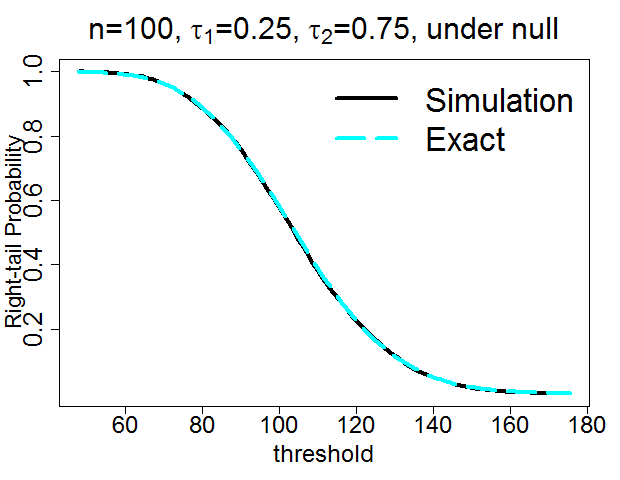}
\end{tabular}%
\caption{The right-tail distribution curve of $W(\tau_1, \tau_2)$ under $H_0$. Left panel: $(\tau_1,\tau_2)=(0.05,0.05)$; Right panel: $(\tau_1,\tau_2)=(0.25,0.75)$. Simulation: Curve obtained by $10^4$ simulations; Exact: by formula (\ref{equ.exactPvalue}).}
\label{fig. null}
\end{figure}

Finally, the survival function of $W$ is given by
\begin{equation}
\begin{gathered}
P(W\geq w)=(1-\tau_1)^{n}I_{\{w\leq 0\}}+\sum_{k=1}^n\binom{n}{k}\tau_1^k(1-\tau_1)^{n-k}e^{-w/2}\left(\frac{\tau_2}{\tau_1}\right)^k\sum_{j=0}^{k-1}\frac{[w/2+k\log(\tau_1/\tau_2)]^j}{j!}\\ 
=(1-\tau_1)^{n}I_{\{w\leq 0\}}+e^{-w/2}\sum_{k=1}^n\sum_{j=0}^{k-1}\binom{n}{k}\tau_2^k(1-\tau_1)^{n-k}\frac{[w+2k\log(\tau_1/\tau_2)]^j}{(2j)!!}.
\label{equ.exactPvalue}
\end{gathered}
\end{equation} 
Note that the formula is not continuous in the first term because of the truncation at $\tau_1$. 
Also as a special case, for the soft-thresholding statistic with $\tau_1=\tau_2=\tau$, we have 
\begin{equation*}
P(W_s\geq w)=(1-\tau)^{n}I_{\{w\leq 0\}}+e^{-w/2}\sum_{k=1}^n\sum_{j=0}^{k-1}\binom{n}{k}\tau^k(1-\tau)^{n-k}\frac{w^j}{(2j)!!}.
\end{equation*} 
For given $\tau_1$ and $\tau_2$, this $p$-value calculation is exact. As evidenced by simulations, Figure \ref{fig. null} shows that formula (\ref{equ.exactPvalue}) provides a perfect null distribution curve for the TFisher family $W$ in (\ref{equ.W}). 

\subsection{Calculation for Omnibus Test} 

\begin{figure}[h] \centering%
\begin{tabular}{l}
\includegraphics[width=3in]{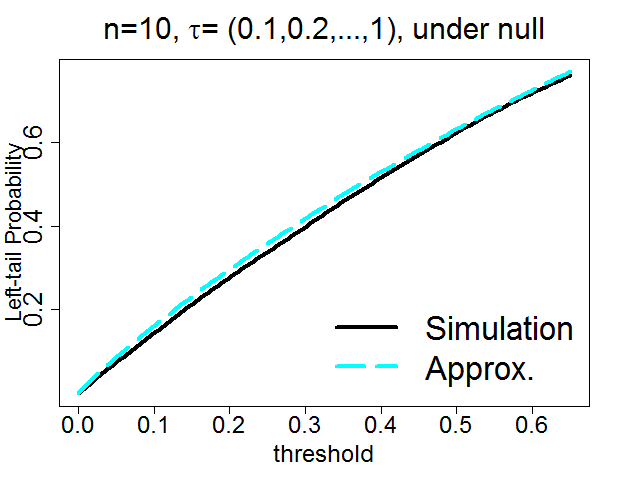}
\includegraphics[width=3in]{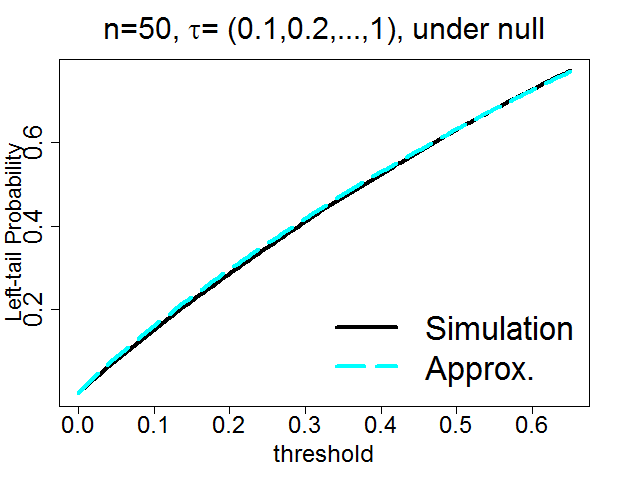}
\end{tabular}%
\caption{The left-tail null distribution of $W_o$ over $\tau_{1j}=\tau_{2j}=\tau_j \in \{0.1, 0.2, ..., 1\}$. Simulation: curve obtained by $10^4$ simulations; Approx.: by calculation in (\ref{equ.exactPvalue}).}
\label{fig. nullomnibus}
\end{figure}

For the omnibus test oTFisher in (\ref{equ.W_o}), noting that $G_j$ is monotone, we have
\begin{equation}
P(\min_j G_j(W_j)>t)=P(W_j(P_1, ..., P_n) < w_j, j=1,...,m),
\end{equation}
where for each $j$ and given $(\tau_{1j}, \tau_{2j})$, the exact value of  $w_j \equiv G_j^{-1}(t)$ can be calculated by (\ref{equ.exactPvalue}). These $W_j$'s are functions of the same set of input $p$-values, and therefore they are dependent among each other.  Fortunately, since $W_j=\sum_{i=1}^n-2\log(P_i/\tau_{2j})I_{(P_i<\tau_{1j})}$, by the Central Limit Theorem (CLT), the statistics $(W_1,...,W_m)$ follow asymptotically the multivariate normal (MVN) distribution with mean vector $\mu = (\mu_1, ..., \mu_m) $ and covariance matrix $\Sigma$, where
\begin{equation}
\begin{gathered}
\mu_j = E(W_j) = 2n\tau_{1j}(1+\log(\tau_{2j}/\tau_{1j})), \text{ and }\\
\Sigma_{jk} = {\rm Cov}(W_j,W_k)  \\
= 4n\tau_{1jk} + 4n\left[\tau_{1jk}(1+\log(\frac{\tau_{2j}}{\tau_{1jk}}))(1+\log(\frac{\tau_{2k}}{\tau_{1jk}}))-\tau_{1j}\tau_{1k}(1+\log(\frac{\tau_{2j}}{\tau_{1j}}))(1+\log(\frac{\tau_{2k}}{\tau_{1k}}))\right],
\end{gathered}
\label{equ.W_mu_Sigma}
\end{equation}
where $\tau_{1jk}=\min\{\tau_{1j}, \tau_{1k}\}$. 
Note that under the special case of the soft-thresholding with $\tau_{1j}=\tau_{2j}=\tau_j$, the two formulas can be readily simplified (assuming $\tau_j\leq\tau_k$) as
$$
\mu_j = 2n\tau_j, \quad{}\Sigma_{jk} = 4n\tau_j\left[2-\tau_{k}+\log(\frac{\tau_{k}}{\tau_j})\right].
$$
Thus we can approximate the $p$-value of oTFisher by the asymptotic distribution of $W_j$'s
\begin{equation}
P(\min_j G_j(W_j)>w_o)\approx P(W'_j<w_j, j=1,...,m),
\end{equation}
where $(W'_j)\sim {\rm MVN}(\mu,\Sigma)$, and $\mu$ and $\Sigma$ are given in (\ref{equ.W_mu_Sigma}). The multivariate normal probabilities can be efficiently computed, e.g., by \cite{genz1992numerical}. Figure \ref{fig. nullomnibus} shows the left-tail probability of $W_o$, which corresponds to the $p$-value because a smaller $W_o$ indicates a stronger evidence against the null. The figure shows that the calculation method is accurate even for small $n$, and the accuracy improves as $n$ increases. The calculation is slightly conservative, which guarantees that the type I error rate will be sufficiently controlled in real  applications. 

%%%========================================

\section{TFisher Distribution Under General $H_1$}\label{Sect_Calcu_Distn_Alter}

In this section we provide a methodology for calculating the distribution of TFisher in (\ref{equ.W}) under the general $H_0$ and $H_1$ in (\ref{equ.GeneralHypo}), and thus the statistical power. Even though the calculation is derived asymptotically, it possesses a high accuracy for small to moderate $n$.  

For any given CDF $F_0$ or $F_1$ in (\ref{equ.GeneralHypo}), we define a monotone transformation function on $[0, 1]$: 
\begin{equation}
	D(x) = \left\{
	\begin{array}{l l}
		x & \quad \text{under $H_0 : F_0$}, \\\
		\bar{F}_1(\bar{F}_0^{-1}(x)) & \quad \text{under $H_1 : F_1 \neq F_0$}.
	\end{array} \right.
\label{equ.D}
\end{equation}
For any random $p$-value $P_i$ in (\ref{equ.P}), we have $D(P_i) \sim$ Uniform$[0, 1]$ under either $H_0$ or $H_1$. Furthermore, we define function
\begin{equation}
\label{equ.deltax}
\delta(x) = D(x) -x,
\end{equation}
which provides a metric for the difference between $H_0$ and $H_1$. For example, for any level $\alpha$ test, $\delta(\alpha)$ represents the difference between the statistical power and the size. For any random $p$-value $P$, $\delta(P)$ measures a stochastic difference between the $p$-value distribution under $H_0$ versus that under $H_1$. 

The TFisher statistic can be written as 
\begin{equation}
W=\sum_{i=1}^{n}-2\log \left( \frac{P_{i}}{\tau_2 }\right) I_{(P_{i}\leq \tau_1 )}
= \sum_{i=1}^{n} Y_i,
\label{equ.W_by_Y}
\end{equation}
where $Y_i \equiv -2\log \left( \frac{D^{-1}(U_{i})}{\tau_2 }\right) I_{(D^{-1}(U_{i})\leq \tau_1 )}$, and $U_i = D(P_i)$ are i.i.d. Uniform$[0, 1]$.

For arbitrary $F_0$ and $F_1$, the $D$ function could be complicated and exact calculation could be difficult. Here we propose an asymptotic approximation for the distribution of $W$ under $H_1$. Note that since $W$ is the sum of i.i.d. random variables, it is asymptotically normal  by the CLT. However, for small to moderate $n$ and for small truncation parameter $\tau_1$, the normal approximation is not very accurate. Here we use a three-parameter $(\xi,\omega,\alpha)$ skew normal distribution (SN) to accommodate the departure from normality  \citep{azzalini1985class}. 
Specifically, we approximate $W$ by 
\begin{equation*}
W \overset{D}{\approx} {\rm SN}(\xi,\omega,\alpha),
\end{equation*}
where the probability density function of SN is 
\begin{equation*}
f(x)=\frac{2}{\omega}\phi\left(\frac{x-\xi}{\omega}\right)\Phi\left(\alpha\frac{x-\xi}{\omega}\right),
\end{equation*}
with $\phi$ and $\Phi$ being the probability density function and the CDF of $N(0,1)$, respectively. The parameters $(\xi,\omega,\alpha)$ are obtained by solving the equations of the first three moments:
\begin{align*}
\xi &= \mu - \left(\frac{2\mu_3}{4-\pi}\right)^{1/3},\\
\omega &= \sqrt{\sigma^2+\left(\frac{2\mu_3}{4-\pi}\right)^{2/3}},\\
\alpha &= {\rm sgn}(\mu_3)\sqrt{\frac{\pi(2\mu_3)^{2/3}}{2\sigma^2(4-\pi)^{2/3}+(2-\pi)(2\mu_3)^{2/3}}},
\end{align*}
where 
\begin{align*}
\mu &= E(W) = nE(Y_1),\\
\sigma^2 &= {\rm Var}(W) = n[E(Y_1^2) - E^2(Y_1)], \\
\mu_3 &= E(W - E(W))^3 = n[E(Y_1 - E(Y_1))^3],
\end{align*}
with 
 \begin{align*}
   EY_1^k &= \int_{0}^{D(\tau_1)} \left(-2\log \left( \frac{D^{-1}(u)}{\tau_2 }\right) \right)^k du \text{, } \quad k=1,2,3.
\end{align*}
 
\begin{figure}[h] \centering%
\begin{tabular}{l}
\includegraphics[width=3in]{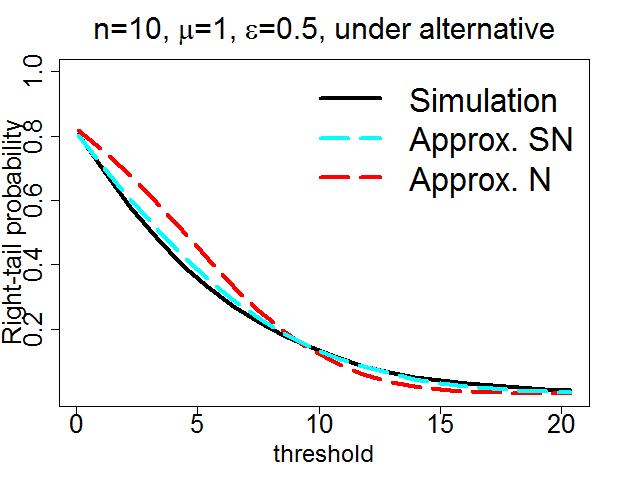}
\includegraphics[width=3in]{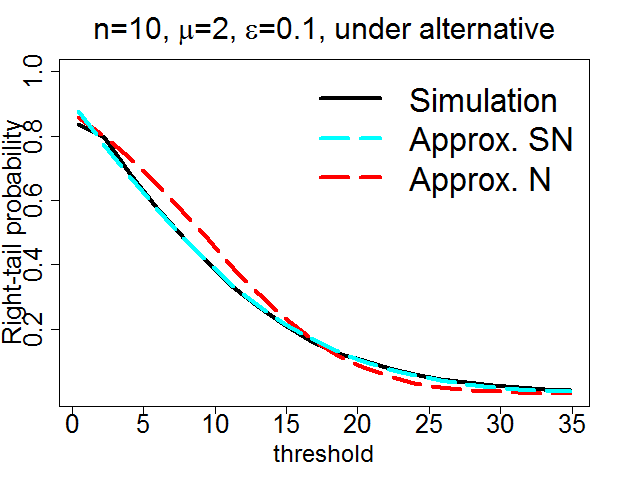}
\end{tabular}%
\caption{The right-tail distribution of $W$ under the alternative hypotheses of Gaussian mixture in (\ref{equ.GaussianHypo}). Left panel: $(\tau_1,\tau_2)=(0.05,0.05)$; right panel: $(0.10,0.25)$. Simulation: curve obtained by $10^4$ simulations; Approx. SN: by the skew-normal approximation; Approx. N: by the normal approximation.}
\label{fig. alter}
\end{figure}

As shown by Figure \ref{fig. alter}, the SN approximation for calculating statistical power is accurate even for small $n$ and $\tau_1$. We have also studied other distribution-approximation techniques including the generalized normal distribution \citep{nadarajah2005generalized, varanasi1989parametric}, the first- and second-order Edgeworth expansions \citep{dasgupta2008asymptotic}, Saddle point approximation \citep{daniels1954saddlepoint, lugannani1980saddle}, etc. Based on our simulation results (not reported in this paper to save space), we note that the SN approximation provides a better accuracy for calculating the power of TFisher with small $\tau_1$ under small $n$.   
  
%%%========================================

\section{Asymptotic Optimality for Signal Detection}\label{Sect_Asymp_Optimality}

In this section, we study the asymptotic performance and optimality within the TFisher family in (\ref{equ.W}). The subscript $n$ is explicitly added to indicate that the asymptotics is driven by $n\to\infty$. %We first study under a classic theory of Bahadur Efficiency. Then, we create a new theory of Asymptotic Power Efficiency, which is more sophisticated and can better explain the properties of these tests in real data analysis. 
Overall, both studies of BE and APE consistently conclude that the soft-thresholding with $\tau_1=\tau_2$ is optimal or close to optimal in a broad space of the signal parameters $(\epsilon,\mu)$, whereas Fisher's method (i.e., no truncation) or TPM (i.e., the hard-thresholding) are not. The functional relationship between optimal $(\tau^*_1,\tau^*_2)$ and $(\epsilon,\mu)$ by APE better reflects the patterns of statistical power than that by BE in real data analysis. 

\subsection{Properties Based on Bahadur Efficiency}

BE was first introduced by \cite{bahadur1960stochastic} to study the large sample property of test statistics. Consider a test $T_n=T(X_1,...,X_n)$, where $X_1,...,X_n$ are random samples. Denote $L_n(t) = P_{H_0}(T_n > t)$ as the survival function of $T_n$ under $H_0$, and $L_n(t | \theta)$ as the survival function under $H_1$. Under $H_1$, if 
\begin{equation}
\lim_{n\to \infty} -\frac{2}{n}\log L_n(T_n|\theta) = c_T(\theta) \in (0, \infty),
\label{equ.c_T}
\end{equation}
we call the constant $c_T(\theta)$ the Bahadur Efficiency (BE, or Bahadur exact slope) of $T_n$ \citep{nikitin1995asymptotic}. Since $L_n(T_n|\theta)$ is actually the $p$-value under $H_1$, $c_T(\theta)$ suggests how quickly the $p$-value decays to zero. Thus, BE indicates how much the null and alternative distributions of $T_n$ are separated  in an asymptotic sense. It is also related to the minimal sample size $n$ that is necessary for the test to reach a given statistical power at a given significance level \citep{bahadur1967rates}. If another test $T^\prime$ has $c_{T^\prime}(\theta) > c_T(\theta)$, $T_n^\prime$ is said to be Bahadur asymptotically more efficient than $T_n$.  Here, for the signal detection problem defined in (\ref{equ.GaussianHypo}), the parameter $\theta$ is a vector $(\epsilon, \mu)$.

Note that under the hypothesis settings in (\ref{equ.GeneralHypo}) and (\ref{equ.GaussianHypo}), the input statistics $X_1, ..., X_n$ can be regarded as the input samples for the $p$-value combination tests, e.g., in (\ref{equ.W}). Thus the number $n$ of tests to be combined can be regarded as the sample size $n$ in the Bahadur asymptotics given in (\ref{equ.c_T}). This setting is similar as some BE studies for $p$-value combination methods (e.g., \cite{abu2003exact}), but are different from the others where the input statistics $X_i$ are related to the sample size (e.g., \cite{littell1971asymptotic,littell1973asymptotic}).

To calculate $c_T(\theta)$, one can apply a composition method (cf. Theorem 1.2.2 in \cite{nikitin1995asymptotic}).  
Specifically, if (i) $T_n \overset{P}{\rightarrow} g(\theta)$ under $H_1$, and (ii) the tail property of $p$-value under $H_0$ satisfies $\lim_{n\to \infty} -\frac{2}{n}\log L_n(t) = f(t)$, where $f(t)$ is continuous on an open interval $I$ and $g(\theta) \in I$ for all $\theta$ under $H_1$, then $c_T(\theta) = f(g(\theta))$. Note that the convergency $T_n \overset{P}{\rightarrow} g(\theta)$ under $H_1$ implies that the variance of $T_n$ will converge to 0 under $H_1$. Thus BE contains the variance information only under $H_0$. We make this important property as a remark. 
\begin{rem}
Bahadur efficiency does not incorporate the information on the variance of the statistic under $H_1$. 
\label{Remark.BE}
\end{rem}
 
Now we calculate the BE of any TFisher statistic $W_n(\tau_1, \tau_2)$ in (\ref{equ.W}). Considering an equivalent test statistic $T_n = W_n/n$ and following (\ref{equ.W_by_Y}) and the Law of Large Numbers, under $H_1$ we have
\begin{align*}
\frac{W_n}{n} \overset{P}{\to} E_1 = E_1(Y_i) 
%=\int_{0}^{D(\tau_1)}-\log \left( \frac{D^{-1}(u)}{\tau_2 }\right)du 
= \int_{0}^{\tau_1}-\log \left( \frac{u}{\tau_2 }\right)D^\prime(u)du.
\end{align*}
Note that,
\begin{align*}
P(W_n / n > t) =P(\frac{ \frac{1}{n} \sum_i^n Y_i - E_0}{\sqrt{V_0/n}}>\frac{t - E_0}{\sqrt{V_0/n}}),
%&\to\bar{\Phi}(\sqrt{n}\frac{t-E_0Z}{\sqrt{Var_0Z}})
\end{align*}
where $E_0$ and $V_0$ denote the mean and variance of $Y_i$ under $H_0$, respectively:
%\begin{align*}
%E_0 &= E_0(Y_i) = \tau_1(1-\log\tau_1+\log\tau_2),\\
%V_0 &= Var_0(Y_i) = \tau_1(1+(1-\tau_1)(1-\log\tau_1+\log\tau_2)^2).
%\end{align*}
\begin{equation}
\label{equ.E0V0}
\begin{gathered}
E_0 = E_{H_0}(Y_i) = \tau_1(1-\log\tau_1+\log\tau_2),\\
V_0 = {\rm Var}_{H_0}(Y_i) = \tau_1(1+(1-\tau_1)(1-\log\tau_1+\log\tau_2)^2).
\end{gathered}
\end{equation}
Consider the statistic under $H_0$, by the CLT and Mill's ratio, we have 
\begin{align*}
\lim_{n\to \infty} - \frac{2}{n}\log P(W_n/n>t) %\to \bar{\Phi}(\sqrt{n}\frac{t-E_0Z}{\sqrt{Var_0Z}}) 
= \frac{(t-E_0)^2}{V_0}.
\end{align*}
Thus the BE of $W_n$ is 
\begin{equation}
\label{equ.c_theta}
c(\epsilon, \mu; \tau_1, \tau_2) = \frac{(E_1-E_0)^2}{V_0}=\frac{\Delta^2}{V_0}.
\end{equation}
The signal parameters $\epsilon$ and $\mu$ are involved through the $D^\prime(u)$ function in the expression of $E_1$. The formula does not contain information on the variance of the statistic under $H_1$, as stated in Remark \ref{Remark.BE}.

%Explicitly, $c(\epsilon, \mu; \tau_1, \tau_2)$ can be re-written as
%\begin{align*}
%c(\epsilon, \mu; \tau_1, \tau_2)=\frac{[\int_0^{\tau_1}-2\log(u)(D^\prime(u)-1)du-2\log(\tau_2)(D(\tau_1)-\tau_1)]^2}{4\tau_1(1+(1-\tau_1)(1-\log(\tau_1)+\log(\tau_2))^2)}
%\end{align*}

The BE-optimal  $\tau_1$ and $\tau_2$ are the ones that maximize $c(\epsilon, \mu; \tau_1, \tau_2)$. Under the general hypotheses in (\ref{equ.GeneralHypo}), based on the metric $\delta(x)$ for the difference between $H_0$ and $H_1$ defined in (\ref{equ.deltax}), Lemma \ref{Thm.C_StaPoint} gives a loose condition for the soft-thresholding being ``first-order optimal" in the sense that it reaches the stationary point of maximization. It means that in a very general case of arbitrary $H_1$, the soft-thresholding with $\tau_1=\tau_2$ may provide a promising choice for construction of a powerful test. 
\begin{lemma}
\label{Thm.C_StaPoint}
Consider TFisher statistics $W_n(\tau_1, \tau_2)$ in (\ref{equ.W}) under the general hypotheses in (\ref{equ.GeneralHypo}). With $\delta(x)$ in (\ref{equ.deltax}), if $\tau^*$ is the solution of equation 
\begin{equation}
\int_0^{x}\log (u) d \delta(u) = \delta(x)\left(\log(x) - \frac{2-x}{1-x}\right),
\label{equ.BE.stationaryPoint}
\end{equation}
then the soft-thresholding with $\tau_1=\tau_2=\tau^*$ satisfies the first-order conditions for maximizing $c(\epsilon, \mu; \tau_1, \tau_2)$ in (\ref{equ.c_theta}). 
\end{lemma}

Equation (\ref{equ.BE.stationaryPoint}) can be easily checked, and is often satisfied in broad cases, e.g., the signal detection problem defined by the Gaussian mixture model in (\ref{equ.GaussianHypo}). However, before getting the specific maximizers $\tau_1^*$ and $\tau_2^*$ of BE, we study their first-order property in a more general case than the Gaussian: hypotheses based on a general mixture model with arbitrary continuous CDFs $G_0$ and $G_1$: 
\begin{equation}
H_0 : X_i \overset{{\rm i.i.d.}}{\sim} G_0 \quad \text{ vs. } \quad H_1 : X_i \overset{{\rm i.i.d.}}{\sim} \epsilon G_1 + (1-\epsilon)G_0,
\label{equ.mixtureModel}
\end{equation}
where the proportion $\epsilon \in (0, 1)$ can be considered as the signal proportion. Lemma \ref{Lem.mixedModel_noEpsi} gives a somewhat surprising result that the maximizers $\tau_1^*$ and $\tau_2^*$ of BE are irrelevant to $\epsilon$. 

\begin{lemma}
\label{Lem.mixedModel_noEpsi}
Consider TFisher statistics $W_n(\tau_1, \tau_2)$ in (\ref{equ.W}) under the hypotheses of mixture model in (\ref{equ.mixtureModel}), the maximizers $\tau_1^{*}$ and $\tau_2^{*}$ of $c(\epsilon, \mu; \tau_1, \tau_2)$ do not depend on $\epsilon$.
\end{lemma}

The result of Lemma \ref{Lem.mixedModel_noEpsi} becomes not so surprising if we consider the limitation of BE as stated in Remark \ref{Remark.BE}. In particular, the denominator $V_0$ of BE in (\ref{equ.c_theta}) represents the variation of the test under $H_0$, which is irrelevant to $\epsilon$. BE is related to $H_1$ only through the difference of the means $E_1-E_0$, which is proportional to $\epsilon$ in the same way no matter what $\tau_1$ and $\tau_2$ are.

For the signal detection problem defined by the Gaussian mixture model in (\ref{equ.GaussianHypo}), Theorem \ref{Thm.C_LocalMax} gives a sufficient condition that guarantees soft-thresholding will reach a local maximum. 

\begin{theorem}
\label{Thm.C_LocalMax}
Consider TFisher statistics $W_n(\tau_1, \tau_2)$ in (\ref{equ.W}) under the signal detection problem in (\ref{equ.GaussianHypo}). Follow the same notations in Lemma \ref{Thm.C_StaPoint}. It can be shown that the solution $\tau^*$  of equation (\ref{equ.BE.stationaryPoint}) exists for any $\epsilon \in (0, 1)$ and $\mu>0.85$. Furthermore, if $\tau^*$ also satisfies the condition
\[
\frac{\delta(\tau^*)}{\delta^\prime(\tau^*)} = \frac{1-\tau^*-\Phi(\Phi^{-1}(1-\tau^*)-\mu)}{e^{\mu\Phi^{-1}(1-\tau^*)-\mu^2/2}-1} >2-\tau^*, 
\]
then the soft-thresolding with $\tau_1=\tau_2=\tau^*$ guarantees a local maximum of $c(\epsilon, \mu; \tau_1, \tau_2)$ in (\ref{equ.c_theta}). In particular, $\tau^*>\bar{\Phi}(\mu/2)$ satisfies the above condition. 
\end{theorem}

\begin{figure}[h] \centering%
\begin{tabular}{l}
\includegraphics[width=2in]{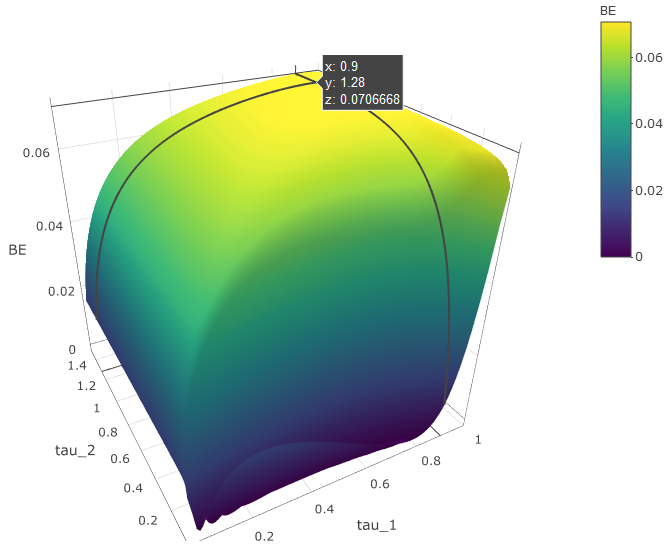}
\includegraphics[width=2in]{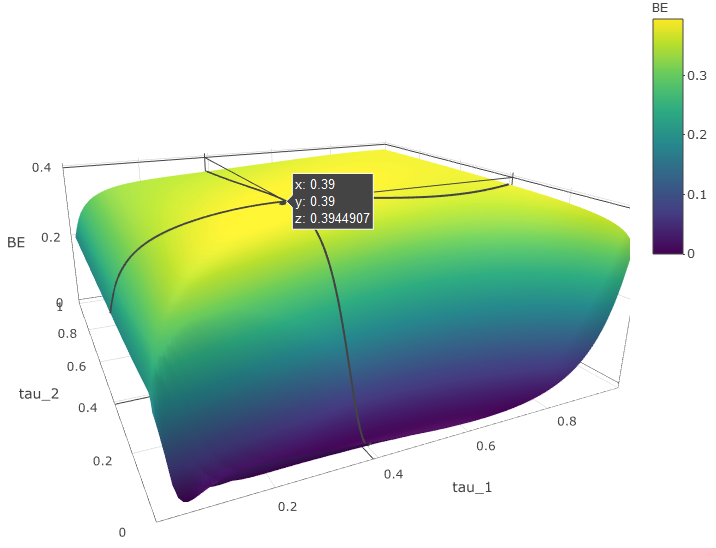}
\includegraphics[width=2in]{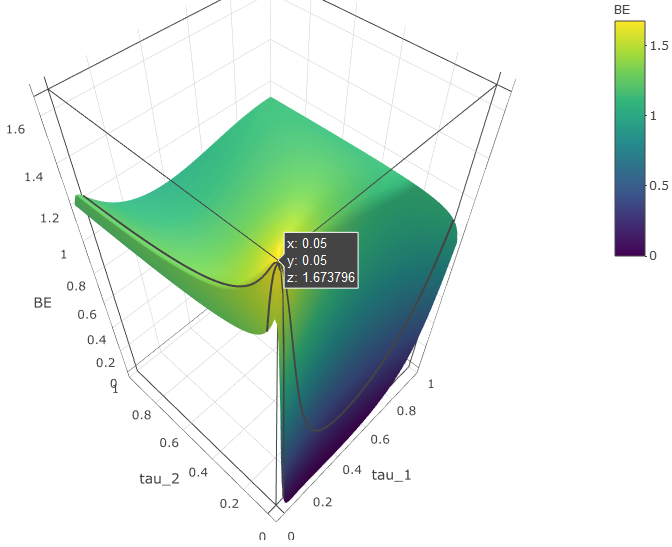}
\end{tabular}%
\caption{3D surface of BE $c(\epsilon, \mu; \tau_1, \tau_2)$ over $\tau_1$ and $\tau_2$. $\epsilon=0.5$.  Left panel: $\mu=0.5$, the maximizers $\tau^*_1= 0.9, \tau^*_2 = 1.28$ and the global maximum $c^*=0.071$; Middle: $\mu=1$, $\tau^*_1= \tau^*_2 = 0.39$ and $c^*=0.394$; Right: $\mu=1.5$, $\tau^*_1= \tau^*_2 = 0.05$ and $c^*=1.674$. }
\label{fig.ctheta_3D}
\end{figure}

Theorem \ref{Thm.C_LocalMax} illustrates that for the signal detection problem, if $\mu$ is not too small, the optimal $\tau^*$ can be calculated. The theorem does not guarantee the maximum is unique. However, since we have the closed form of BE in (\ref{equ.c_theta}), we can always study its properties numerically. Fixing $\epsilon=0.5$, for $\mu = 0.5$, $1$, and $1.5$, Figure \ref{fig.ctheta_3D} gives the numerical values of $c(\epsilon, \mu; \tau_1, \tau_2)$ over a grid of $\tau_1 \in (0, 1)$ and $\tau_2 \in (0, 3)$ with step size 0.01. It shows that the local maximum is unique. More numerical studies under various setups of $\mu$ and $\epsilon$ also confirm that the maximum is likely unique (results not shown here to save space). 

\begin{figure}[h] \centering%
\begin{tabular}{l}
\includegraphics[width=3in]{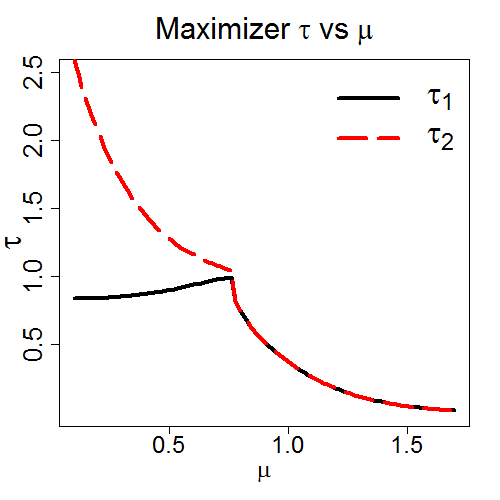}
\includegraphics[width=3in]{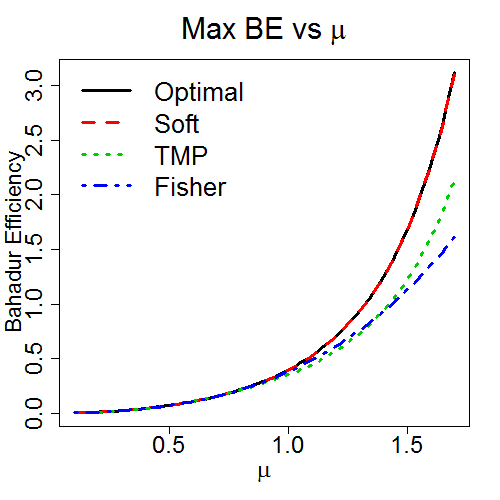}
\end{tabular}%
\caption{BE-optimality over $\mu$ values. Left panel: Global maximizers $\tau^*_1$ and $\tau^*_2$ of BE $c(\epsilon, \mu; \tau_1, \tau_2)$ over $\mu$. Right panel: Maximums of BE over $\mu$. Optimal: Globally maximal BE; Soft: Maximal BE under restriction $\tau_1=\tau_2$; TPM: Maximal BE under restriction $\tau_2=1$; Fisher: BE at $\tau_1= \tau_2=1$. $\epsilon=0.5$.  
}
\label{fig. ctheta_optional}
\end{figure}

To further study the relationship between the maximizers and the maximum of $c(\epsilon, \mu; \tau_1, \tau_2)$, the left panel of Figure \ref{fig. ctheta_optional} shows the values of global maximizers $\tau^*_1, \tau^*_2 $ over $\mu$; the right panel shows the global and restricted maximums. A few observations can be made. First, the soft-thresholding with $\tau^*_1= \tau^*_2 $ is global optimal for maximizing BE when $\mu>0.78$. It indicates that the lower bound for the cut-off of $0.85$ given in Theorem \ref{Thm.C_LocalMax} is pretty tight. Secondly, when $\mu$ is larger than this cutoff, $\tau^*_1=\tau^*_2 =\tau^*$ is a decreasing function of the signal strength $\mu$. That is, the stronger the signals, the more beneficial the truncation method will be. When the signals are weaker, i.e., when $\mu$ is less than the cutoff, the optimal $\tau^*_1, \tau^*_2$ could be different. $\tau^*_1$ is close to 1, but $\tau^*_2$ could be larger than 1. It means that for weak signals, we should not truncate too much, but instead should give a heavier weight to smaller $p$-values through $\tau_2$. Thirdly, even when the soft-thresholding is not the optimal, it still gives a very similar value of $c(\epsilon, \mu; \tau_1, \tau_2)$. That can be seen from the right panel of Figure \ref{fig. ctheta_optional}: when $\mu$ is small, various methods have a similar $c(\epsilon, \mu; \tau_1, \tau_2)$ value, which is close to 0. However, when $\mu$ is large, we note that the optimal soft-thresholding is significantly better than the optimal hard-thresholding (TPM), and both are better than Fisher's method (no truncation). This result means that the difference between soft-thresholding and the global-optimal methods could be practically negligible.

\subsection{Properties Based on Asymptotic Power Efficiency}

BE has a limitation for fully reflecting the statistical power of a given test. Following Remark \ref{Remark.BE} and Lemma \ref{Lem.mixedModel_noEpsi}, for any mixture model in (\ref{equ.mixtureModel}) BE does not reflect the influence of $\epsilon$ to statistical power, which could be not true in real data analysis. In particular, BE in (\ref{equ.c_theta}) is related to $H_1$ only through the difference of the means $E_1-E_0$, but not the variance.  However, in reality a given statistic could have significantly different variations under the null and the alternative. To address this limitation, we develop a new asymptotic metric, called Asymptotic Power Efficiency (APE), which will take such variation difference into consideration. %Therefore, contrasting to BE, APE is a more direct and accurate asymptotics to reflect the patterns of statistical power. The optimality of $\tau_1$ and $\tau_2$ by maximizing APE will be more directly related to statistical power, which is indeed relevant to $\epsilon$.  

%APE is motivated by the direct calculation of statistical power, except under asymptotics. 
APE is defined based on a more direct and accurate asymptotics to reflect the patterns of statistical power. Following equation (\ref{equ.W_by_Y}), under $H_0$, by the CLT we have
\[
P_{H_0}(W_n>nE_0 + z_\alpha\sqrt{nV_0}) \to \alpha, 
\]
where $E_0$ and $V_0$ are defined in (\ref{equ.E0V0}), $z_\alpha$ is the $(1-\alpha)$ quantile of $N(0, 1)$. 
We call $nE_0 + z_\alpha\sqrt{nV_0}$ the level-$\alpha$ asymptotic critical value for $W_n$. 
Accordingly, the asymptotic power is 
\begin{align*}
P_{H_1}\left(W_n>nE_0 + z_\alpha\sqrt{nV_0}\right)=P\left(\frac{W_n-nE_1}{\sqrt{nV_1}}>z_\alpha\sqrt{\frac{V_0}{V_1}}-\sqrt{n}\frac{E_1-E_0}{\sqrt{V_1}}\right),
\end{align*}
where 
\begin{align*}
V_1= E_{H_1}(Y_i) = &[ \int_{0}^{\tau_1}\log^2 (u)D^\prime(u)du - (\int_{0}^{\tau_1}\log (u)D^\prime(u)du)^2 \\
+ &2(D(\tau_1)-1)\log(\tau_2)\int_{0}^{\tau_1}\log (u)D^\prime(u)du +\log^2(\tau_2)D(\tau_1)(1-D(\tau_1))].
\end{align*}
The rescaled critical value 
\begin{equation}
\label{equ.a_theta}
a(\epsilon, \mu; \tau_1, \tau_2) = z_\alpha\sqrt{\frac{V_0}{V_1}}-\sqrt{n}\frac{\Delta}{\sqrt{V_1}}
\end{equation}
is called the APE. Since $\frac{W_n-nE_1}{\sqrt{nV_1}} \rightarrow N(0, 1)$, the smaller the $a(\epsilon, \mu; \tau_1, \tau_2)$, the bigger the asymptotic power, and thus the more ``efficient" a test is. BE and APE are consistent in the sense that the bigger the mean difference $\Delta$, the more efficient a test is. Meanwhile, APE is more sophisticated as it accounts for differences of both the means and the variances under the alternative versus the null. 

When $n$ is large, $a(\epsilon, \mu; \tau_1, \tau_2)$ is dominated by the $\sqrt{n}$ term. We define
\begin{equation}
\label{equ.b_theta}
b(\epsilon, \mu; \tau_1, \tau_2) = \frac{\Delta}{\sqrt{V_1}}
\end{equation}
as another measure for the performance of a statistic, called Asymptotic Power Rate (APR). Note that APR is similar as BE except that the denominator refers to the alternative variance under $H_1$. Since APR is more directly related to statistical power than BE, this formula indicates that the variance of the statistic under the alternative hypothesis could be more relevant to its power than its null variance.  The next theorem indicates that the soft-thresholding method can be a promising candidate in terms of maximizing $b(\epsilon, \mu; \tau_1, \tau_2)$, as long as the signal strength $\mu$ is not too small and the signal proportion $\epsilon$ is not too large. 

\begin{theorem}
\label{Thm.btheta}
Consider TFisher statistics $W_n(\tau_1, \tau_2)$ in (\ref{equ.W}) under signal detection problem in (\ref{equ.GaussianHypo}). When $\mu> 0.85$ and 
\[
\epsilon<h_b(\mu)=\frac{1+\tilde{g}_1(\mu)}{(\tilde{g}_1(\mu))^2-\tilde{g}_1(\mu)-\tilde{g}_2(\mu)},
\]
where $\tilde{g}_k(\mu)=\int_{0}^1\log^k(u)(e^{\mu\Phi^{-1}(1-x)-\mu^2/2}-1)du$, the soft-thresolding with $\tau_1=\tau_2=\tau^*$, for some $\tau^*$, is a stationary point of $b(\epsilon, \mu; \tau_1, \tau_2)$ in (\ref{equ.b_theta}). 
\end{theorem}

Comparing with Theorem~\ref{Thm.C_LocalMax} for BE, Theorem~\ref{Thm.btheta} for APR provides a consistent, yet more comprehensive picture about the optimality domain involving $\epsilon$. Moreover, we give a similar theorem concerning the APE, which further allows the number of tests $n$ and the significance level $\alpha$ to play a role in determining the theoretical boundary for the soft-thresholding to be promising. 

\begin{theorem}
\label{Thm.atheta}
Follow the assumptions and notations in Theorem~\ref{Thm.btheta}. There exists a lower bound $\underline{\mu}^\prime>0$ such that if $\mu>\underline{\mu}^\prime$ and 
\[
\epsilon< h_a(\mu)  =\frac{(1-c_n)[1+\tilde{g}_1(\mu)]+2\tilde{g}_1(\mu)+\tilde{g}_2(\mu)}{(1-c_n)[(\tilde{g}_1(\mu))^2-\tilde{g}_1(\mu)-\tilde{g}_2(\mu)]+2\tilde{g}_1(\mu)+\tilde{g}_2(\mu)},
\]
where $c_n=\sqrt{n}/z_\alpha$, then the soft-thresolding with $\tau_1=\tau_2=\tau^*$, for some $\tau^*$, is a stationary point of $a(\epsilon, \mu; \tau_1, \tau_2)$ in (\ref{equ.a_theta}). 
\end{theorem}

\begin{figure}[H] \centering%
\begin{tabular}{l}
\includegraphics[width=3.5in]{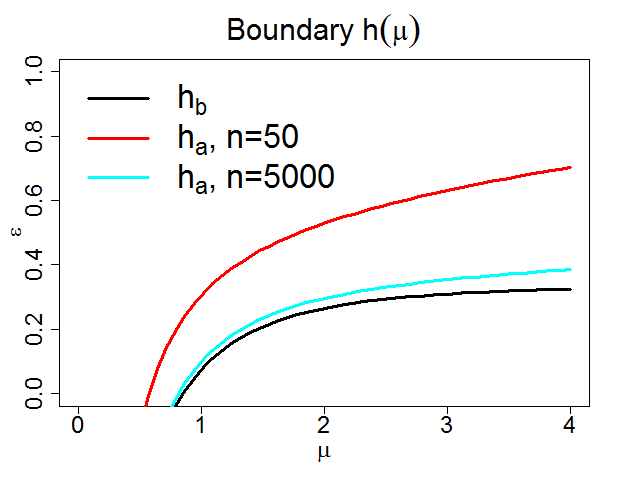}
\end{tabular}%
\caption{The boundaries defined by $h_b(\mu)$ (Theorem \ref{Thm.btheta}, black) and $h_a(\mu)$ (Theorem \ref{Thm.atheta}, $\alpha=0.05$, red: $n = 50$; cyan: $n = 5000$). The soft thresholding $\tau_1=\tau_2=\tau^*$, for some $\tau^*$, satisfies the first order condition of maximizing $b(\epsilon, \mu; \tau_1, \tau_2)$ or $a(\epsilon, \mu; \tau_1, \tau_2)$ for all $(\epsilon,\mu)$ below the corresponding boundary curves.}
\label{fig. boundary}
\end{figure}

Theorems \ref{Thm.btheta} and \ref{Thm.atheta} show that when $\epsilon$ is not too large and $\mu$ is not too small, soft-thresholding is promising. Figure \ref{fig. boundary} shows that when $n$ becomes larger, the theoretical boundary defined by $a(\epsilon, \mu; \tau_1, \tau_2)$ is closer to the boundary defined by $b(\epsilon, \mu; \tau_1, \tau_2)$, as expectd. Under finite $n$, the advantage of soft-thresholding is even more prominent because the curve with $n=50$ covers a bigger parameter space than that of the other two.  

\begin{figure}[H] \centering%
\begin{tabular}{l}
\includegraphics[width=2.5in]{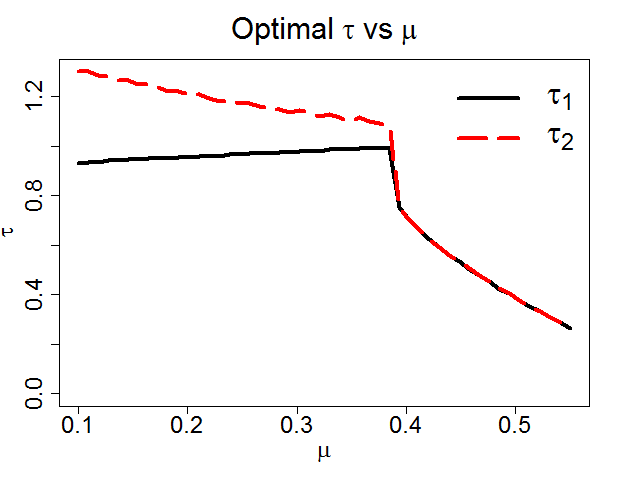}
\includegraphics[width=2.5in]{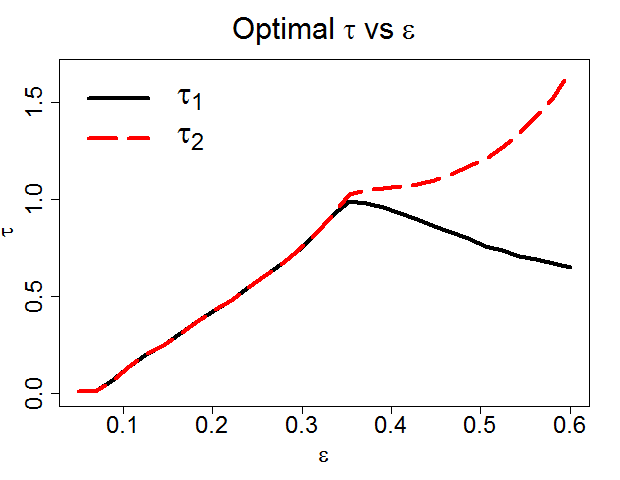}\\
\includegraphics[width=2.5in]{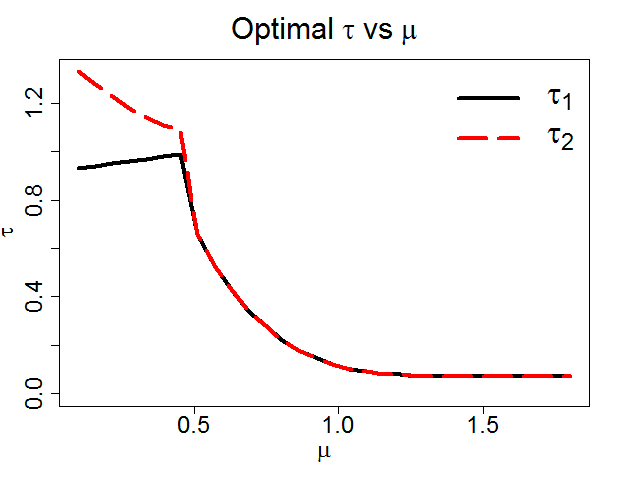}
\includegraphics[width=2.5in]{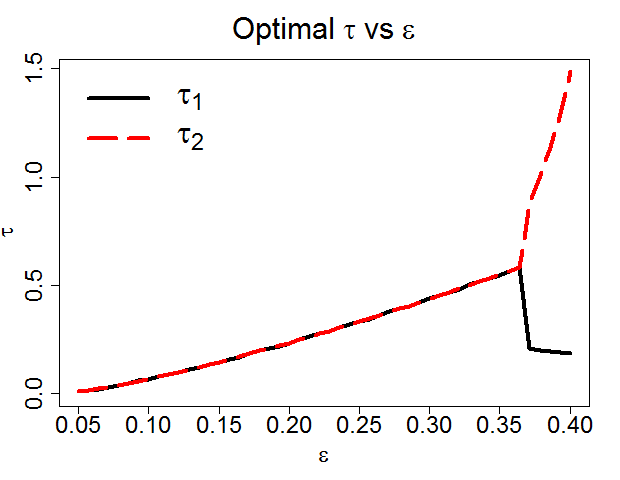}\\
\end{tabular}%
\caption{The global maximizer $(\tau^*_1,\tau^*_2)$ for $a(\epsilon, \mu; \tau_1, \tau_2)$ when $n=50$, $\alpha=0.05$. From top to bottom, left column: $\epsilon = 0.01$ or $0.1$%and $0.25$. 
; right column: $\mu=1$ or $2$. % and $3$
}
\label{fig. atheta}
\end{figure}

We further study numerically the optimal points based on APE.  At $n=50$ and $\alpha=0.05$, the left panels of Figure \ref{fig. atheta} fix $\epsilon$ (row 1: $\epsilon =0.01$; row 2: $\epsilon=0.1$) and plot the maximizer $\tau_1^*, \tau_2^*$ over $\mu$. The pattern is consistent with that for BE in Figure \ref{fig. ctheta_optional}: the soft-thresholding is indeed globally optimal when $\mu$ is large enough, and $\tau^*$ is a decreasing function of $\mu$. 
Moreover, the smaller the $\epsilon$, the smaller the $\mu$ cutoff to guarantee the soft-thresholding being optimal. When $\mu$ is smaller than the cutoff, both $\tau_1^*$ and $\tau_2^*$ could be large, indicating a light truncation and a significance-upscaling weighting for the $p$-values. The right panels of Figure \ref{fig. atheta} fix $\mu$ (row 1: $\mu =1$; row 2: $\mu=2$) and plot the maximizer $\tau_1^*, \tau_2^*$ over $\epsilon$. Consistent with our theorem, the soft-thresholding is indeed globally optimal when $\epsilon$ is not too large (i.e., sparse signals). Such optimal $\tau^*$ is proportional to the signal proportion $\epsilon$. The $\tau^*/\epsilon$ ratio is a decreasing function of  $\mu$, which could be larger or smaller than 1. Thus, the best cutoff $\tau^*$ is not a ``natural" value 0.05 as suggested in literature \citep{zaykin2002truncated}; it is also not simply the signal proportion $\epsilon$. Instead, there is a functional relationship between $\tau^*$ and the signal pattern defined by $\epsilon$ and $\mu$, as is given here. 

\begin{figure}[h] \centering%
\begin{tabular}{l}
\includegraphics[width=2.5in]{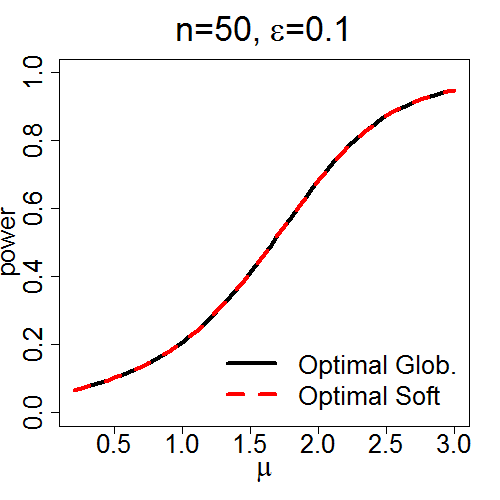}
\includegraphics[width=2.5in]{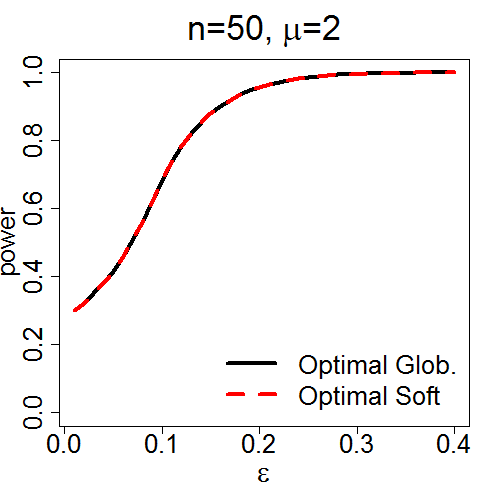}
\end{tabular}%
\caption{Power comparison between globally optimal TFisher statistic (at global maximizers $(\tau^*_1, \tau^*_2)$ of APE) and the optimal soft-thresholding TFisher (at restricted maximizers $\tau_1=\tau_2=\tau^*$ of APE). The number of tests $n=50$, the type I error $\alpha=0.05$. Left: $\epsilon = 0.1$. Right: $\mu=2$.}.
\label{fig. optimalpower}
\end{figure}

When $\epsilon$ is big or when $\mu$ is small, the soft-thresholding may not be optimal based on APE. However, when that happens, the practically meaningful difference is likely small because these areas correspond the true statistical power being close to 0 or 1.  Figure~\ref{fig. optimalpower} shows the comparison of statistical power between the global optimality (with maximizers $(\tau^*_1,\tau^*_2)$ of APE) and the optimal soft-thresholding (under restriction $\tau_1=\tau_2=\tau^*$).  The two power curves match perferctly, even at regions where the soft-thresholding may not be globally optimal in theory. Here the optimization is done by a grid search over $\tau_1 \in \{0.001, 0.002, ..., 1\}$ and $\tau_2 \in \{0.001, 0.002, ..., 10\}$, and statistical power is calculated by the method provided in Section \ref{Sect_Calcu_Distn_Alter}. The result suggests that we may almost always focus on the soft-thresholding in the TFisher family. 

%%%========================================

\section{Statistical Power Comparison For Signal Detection}\label{Sect_Simu}

In this section, we focus on statistical power for the signal detection problem in (\ref{equ.GaussianHypo}). First, we show that our analytical power calculation is accurate when comparing with simulations. Then, we compare the statistical power among different methods and demonstrate their relative performance.

\begin{figure}[h] \centering%
\begin{tabular}{l}
\includegraphics[width=2in]{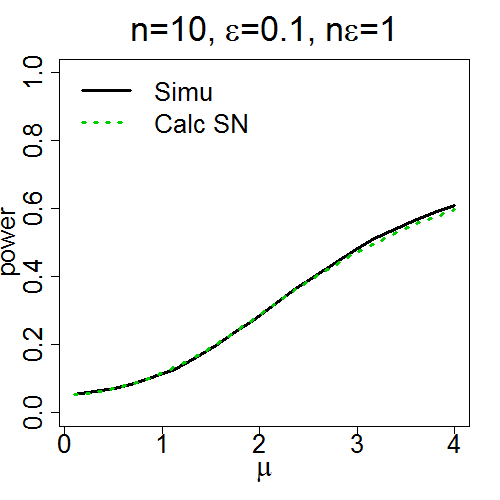}
\includegraphics[width=2in]{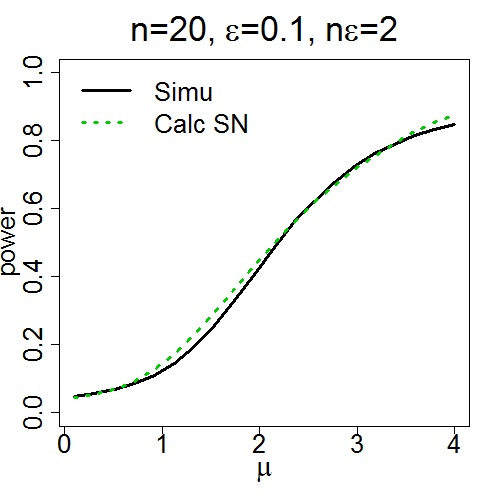}
\includegraphics[width=2in]{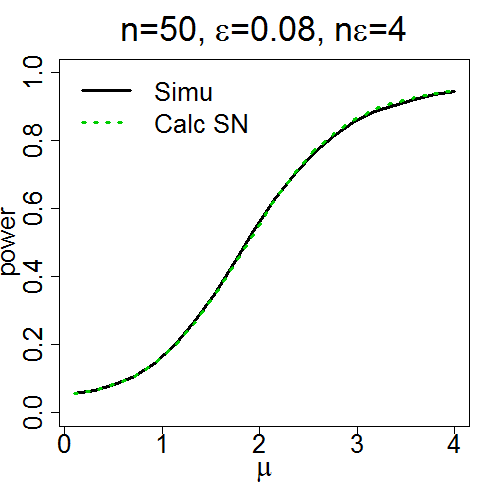}
\end{tabular}%
\caption{The statistical power calculation versus simulation for signal detection. Type I error rate $\alpha=0.05$. Left panel: $\tau_1=0.1, \tau_2=0.5$; Middle: $\tau_1=0.05, \tau_2=0.05$; Right: $\tau_1=0.05, \tau_2=0.25$. Simu: curve by $10^4$ simulations. Calc SN: by calculation.  }
\label{fig. power approx}
\end{figure}

Statistical power calculation combines the calculations for the null distribution (for controlling the type I error) given in Section \ref{Sect_P_Calcu} and for the alternative distribution given in Section \ref{Sect_Calcu_Distn_Alter}. Here we evidence the accuracy of these calculation methods through comparing the statistical power by calculation versus simulation. Figure \ref{fig. power approx} shows that even for relatively small $n$, we have accurate statistical power calculation under various model parameter setups.  

\begin{figure}[H] \centering%
\begin{tabular}{l}
\includegraphics[width=2in]{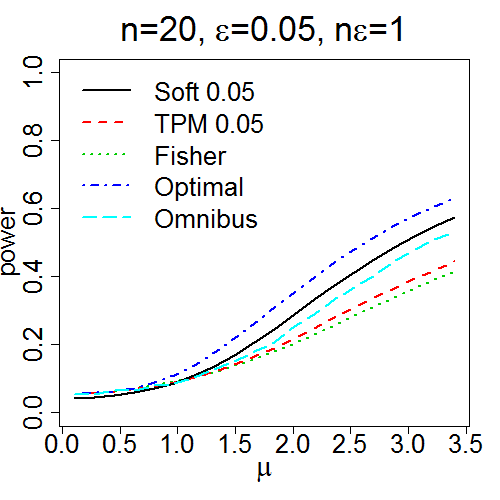}
\includegraphics[width=2in]{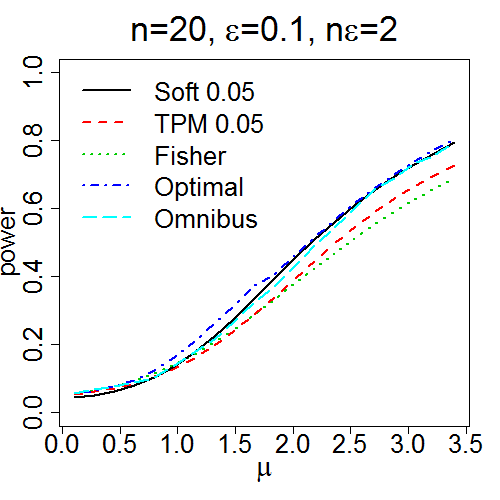}
\includegraphics[width=2in]{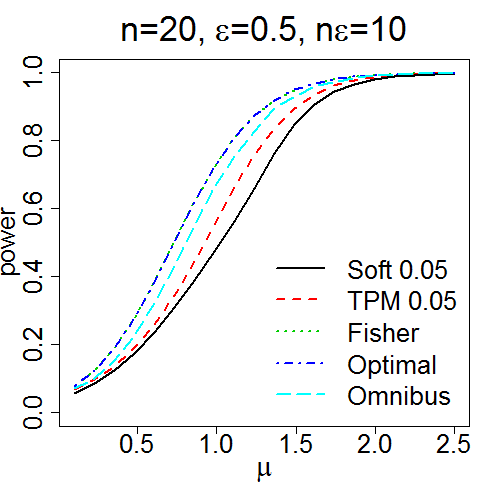}\\
\includegraphics[width=2in]{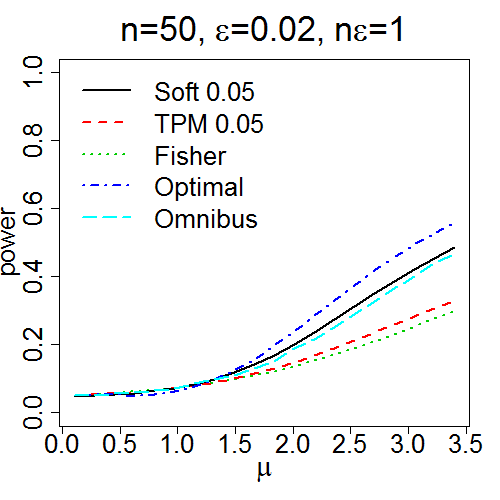}
\includegraphics[width=2in]{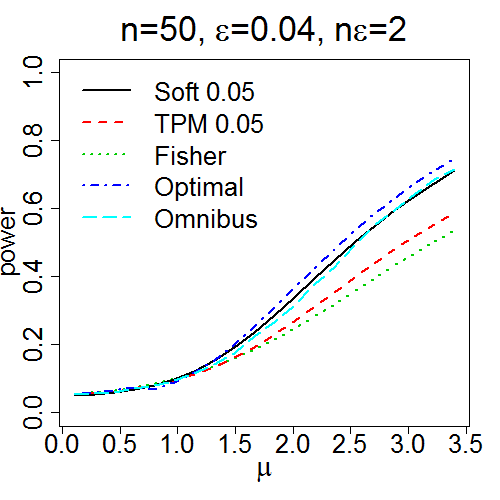}
\includegraphics[width=2in]{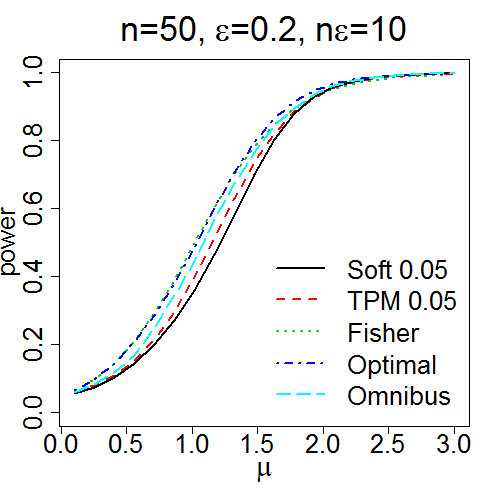}\\
\includegraphics[width=2in]{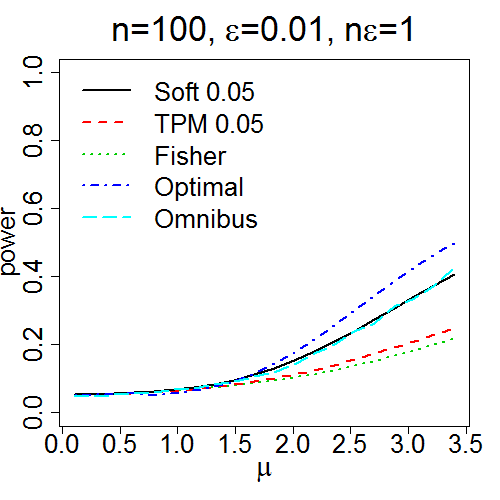}
\includegraphics[width=2in]{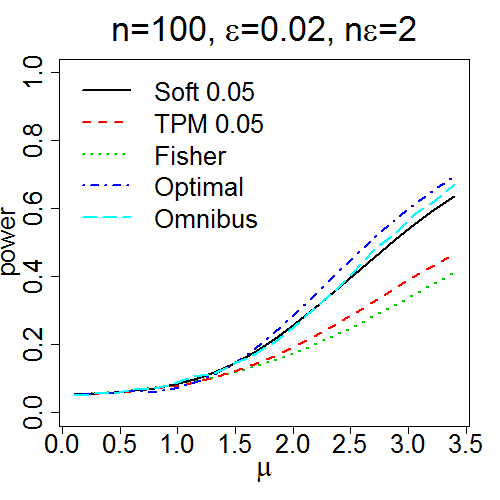}
\includegraphics[width=2in]{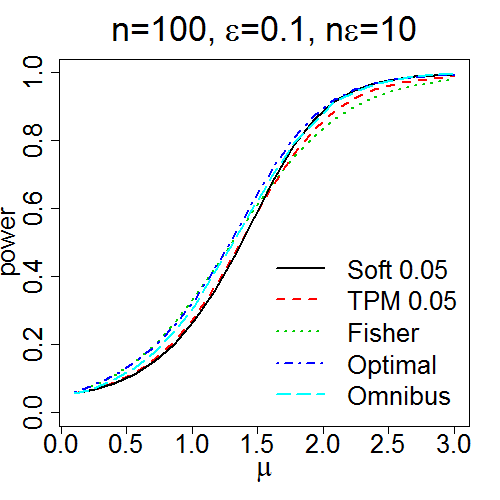}
\end{tabular}%
\caption{The power comparison over signal strength $\mu$. Type I error rate $\alpha=0.05$. Soft 0.05: soft-thresholding at $\tau_1=\tau_2=0.05$; TPM 0.05: hard-thresholding at $\tau_1 = 0.05, \tau_2=1$; Fisher: Fisher's combination at $\tau_1=\tau_2=1$; Optimal: optimal TFisher at maximizers $\tau_1^*,\tau_2^*$ of APE; Omnibus: soft-thresholding oTFisher with adaptive $\tau \in \{0.01,0.05,0.5, 1\}$.}
\label{fig.power_over_mu}
\end{figure}

\begin{figure}[H] \centering%
\begin{tabular}{l}
\includegraphics[width=2in]{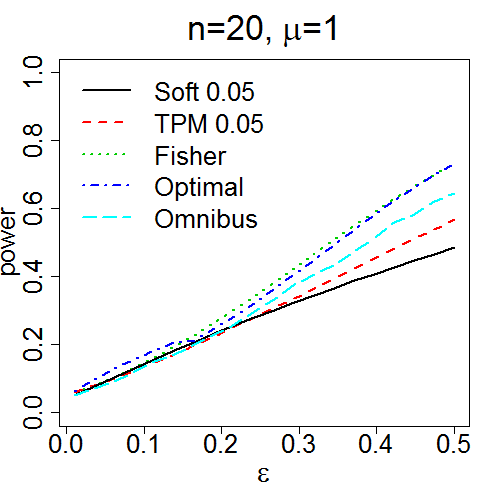}
\includegraphics[width=2in]{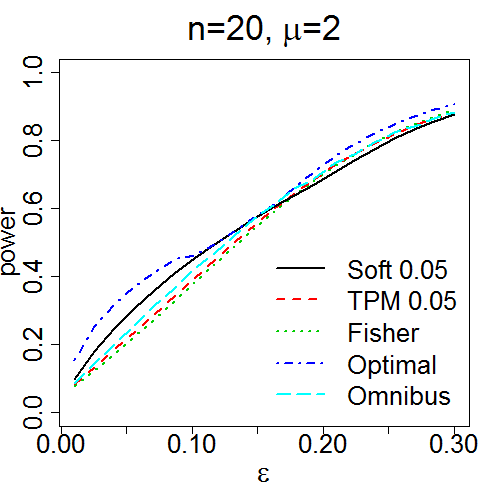}
\includegraphics[width=2in]{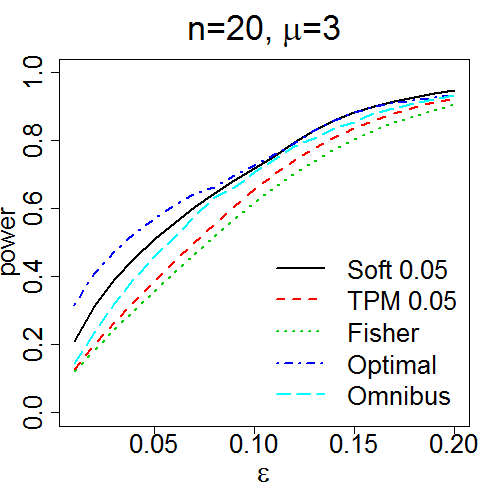}\\
\includegraphics[width=2in]{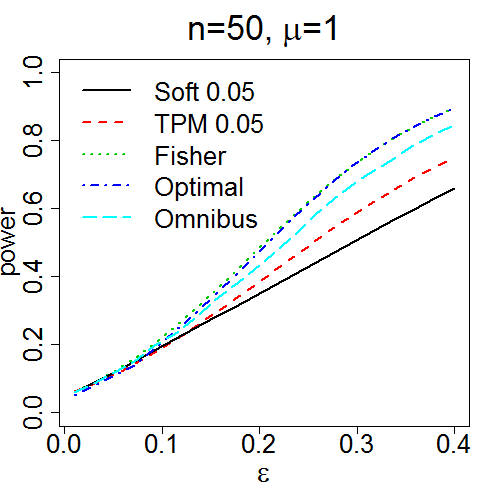}
\includegraphics[width=2in]{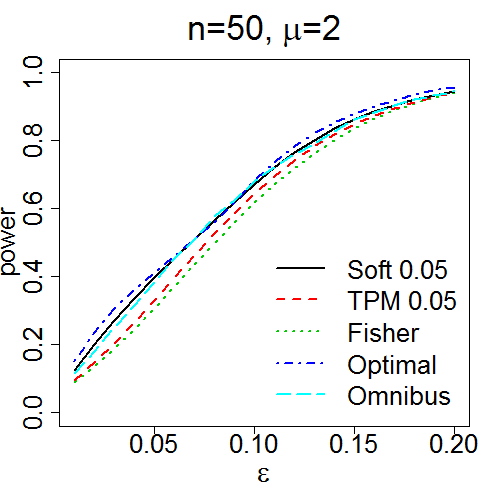}
\includegraphics[width=2in]{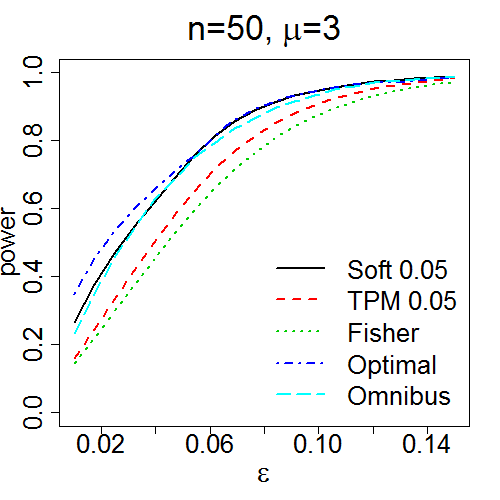}\\
\includegraphics[width=2in]{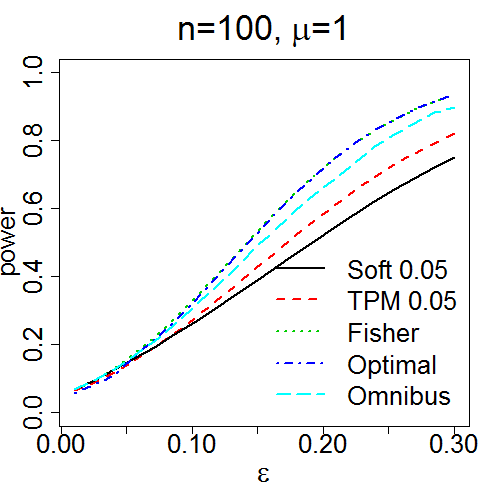}
\includegraphics[width=2in]{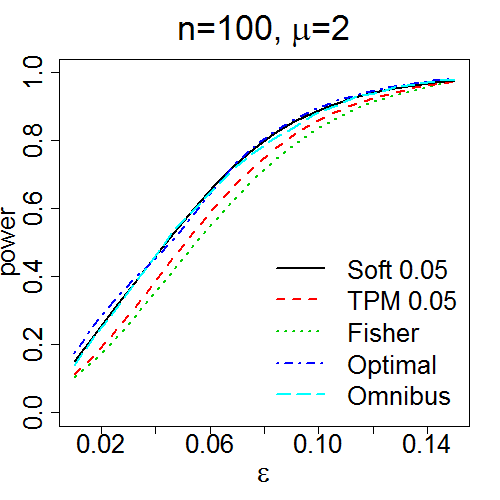}
\includegraphics[width=2in]{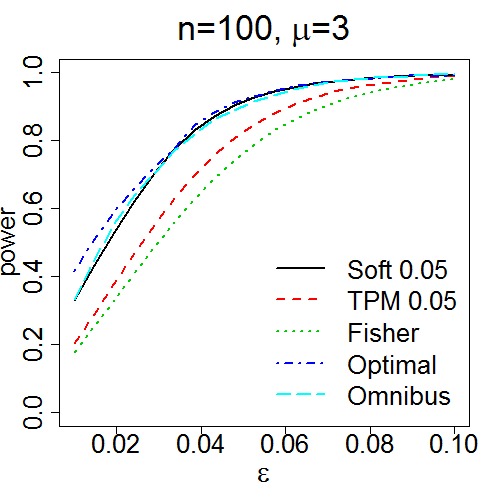}
\end{tabular}%
\caption{The power comparison over signal proportion $\epsilon$. Type I error rate $\alpha=0.05$. Soft 0.05: soft-thresholding at $\tau_1=\tau_2=0.05$; TPM 0.05: hard-thresholding at $\tau_1 = 0.05, \tau_2=1$; Fisher: Fisher's combination at $\tau_1=\tau_2=1$; Optimal: optimal TFisher at maximizers $\tau_1^*,\tau_2^*$ of APE; Omnibus: soft-thresholding oTFisher with adaptive $\tau \in \{0.01,0.05,0.5, 1\}$.}
\label{fig.power_over_eps}
\end{figure}

Next, we compare various methods in the TFisher family: optimal TFisher with global maximizers $\tau_1^*, \tau_2^*$ of APE in (\ref{equ.a_theta}), soft-thresholding with fixed $\tau_1=\tau_2=0.05$, soft-thresholding omnibus test oTFisher with adaptive $\tau_1=\tau_2 \in \{0.01,0.05,0.5, 1\}$, Fisher's method with $\tau_1=\tau_2=1$, and TMP with $\tau_1=0.05$ and $\tau_2=1$. Figure \ref{fig.power_over_mu} illustrates the power over the signal strength $\mu$ at various number $n$ of input $p$-values (by row) and the expected number $n\epsilon$ of signals (by column). Figure \ref{fig.power_over_eps} illustrates the power over signal proportion $\epsilon$ at various $n$ (by row) and the signal strength $\mu$ (by column). Interesting observations can be seen from these two figures. First, with no surprise, the optimal TFisher is always the best over all settings. Actually in most of those cases the optimal TFisher corresponds to the soft-thresholding with $\tau_1^* =\tau_2^*$, and if they are not equal, the power difference is almost always negligible (see Figures \ref{fig. atheta} and \ref{fig. optimalpower}). 
Secondly, the soft-thresholding oTFisher is a relatively robust method over various signal patterns. It is often close to the best, and never be the worst. In fact, its power is often close to the power of the statistic with the parameters it adaptively chooses. For example, if oTFisher chooses $\tau_1=\tau_2=0.05$, it gives a similar but slightly lower power than TFisher with the same parameters. The slight loss of power is possibly due to the variation of the adaptive choice. Thirdly, the soft-thresholding TFisher with fixed $\tau_1=\tau_2=0.05$ has a clear advantage when signals are sparse, i.e., when $\epsilon$ is small. It also has a  clear disadvantage when the signals are dense. The original Fisher's method, which is also a special case of soft-thresholding shows an opposite pattern. Meanwhile, the relative advantage of Soft-0.05 versus Fisher is also related to the signal strength $\mu$. In consistence with the theoretical study of both BE and APE, the larger the $\mu$, the smaller the optimal $\tau^*$ shall be. Such phenomenon is evidenced by panel 3-3 in Figure \ref{fig.power_over_mu} and the panel 1-3 in Figure \ref{fig.power_over_eps}: when $\epsilon$ is relatively big, say around 0.1 and 0.2, Soft-0.05 could still be better than Fisher at large $\mu$. Lastly, the hard-thresholding TMP-0.05 is mostly not among the best. In particular, it has a clear disadvantage to Soft-0.05 for detecting sparse signals. 

Finally we compare the power of three omnibus tests: oTFisher with soft-thresholding, the adaptive TPM (ATPM, hard-thresholding), and the adaptive RTP (ARTP). ARTP was shown to have the highest power among a group of adaptive set-based methods for genetic association testing \citep{Yu2009, su2016adaptive}. The Supplementary Figures \ref{fig.artppower_over_mu} and \ref{fig.artppower_over_eps} 
in the Supplementary Materials illustrate the power of the optimal TFisher and the three omnibus tests under the same settings as Figures  \ref{fig.power_over_mu} and \ref{fig.power_over_eps}, respectively. %$\mu$ and $\epsilon$, respectively. 
The key result is that oTFisher actually dominates both ATPM and ARTP across all settings of signal patterns. 
ARTP could be better than ATPM for sparser and stronger signals, but the opposite is true for denser and weaker signals. 

In summary, the pattern of power comparison well reflects the theoretical study in Section \ref{Sect_Asymp_Optimality}. The soft-thresholding that restricts $\tau_1=\tau_2=\tau$ is the right strategy to reach the optimal statistic in most cases. The optimal $\tau^*$ is related to the signal pattern defined by both parameters $\epsilon, \mu$. If we know the signal pattern, e.g., small $\epsilon$   (especially if $\mu$ is big at the same time), then we should choose a small $\tau$. However, if no such prior information is available in a study, then the soft-thresholding oTFisher with a grid of $\tau$ over small, mediate and large values in $(0, 1)$ will likely be a robust solution.

%%%========================================

\section{ALS Exome-seq Data Analysis}\label{Sect_ExomeSeq}

The $p$-value combination methods have been widely used for genetic association studies, but most of them were based on hard-thresholding, including TPM and RTP methods \citep{Hoh2001, dudbridge2003rank, Yu2009, li2011adaptively, biernacka2011use, hongying2014modified, su2016adaptive}. 
In this section we apply and assess the soft-thresholding TFisher by analyzing a whole exome sequencing data of amyotrophic lateral sclerosis (ALS). ALS is a neurodegenerative disorder resulting from motor neuron death. It is the most common motor neuron disease in adults (Motor Neuron Diseases Fact Sheet, NINDS). ALS is a brutal disease that causes patients to lose muscle strength and coordination even for breathing and swallowing, while leaving their senses of pain unaffected. ALS is uniformly fatal, usually within five years. Genetics plays a critical role in ALS; the heritability is estimated about $61\%$ \citep{smith2014exome}. The identification of ALS genes is foundational in elucidation of disease pathogenesis, development of disease models, and design of targeted therapeutics. Despite numerous advances in ALS gene detection, these genes can explain only a small proportion (about 10\%) of cases \citep{cirulli2015exome}. 

Exome-sequencing data is obtained by the next-generation sequencing technology for sequencing all protein-coding genes in a genome, i.e., the exome. This approach identifies genetic variants that alter protein sequences that may affect diseases. It provides a great balance between the depth of sequencing and the cost comparing with the whole-genome sequencing. Our data comes from the ALS Sequencing Consortium, and the data cleaning and single nucleotide variant (SNV) filtering process follows the same steps as the original study \citep{smith2014exome}. Specifically, we focused on SNVs which occur at highly conserved positions (with positive GERP score \citep{davydov2010identifying}) %"dbNSFP.GERP\_RS $> 0$") 
or which represent stop-gain or stop-loss mutations \citep{liu2016dbnsfp}. %("dbNSFP.aaref='X' OR dbNSFP.aaalt='X'")] 
SNVs that have low genotyping quality (missing rate $< 40\%$) were remove; missing genotypes were also removed. After these filtering steps, the data contained 457 ALS cases and 141 controls, with 105,764 SNVs in 17088 genes. Two non-genetic categorical covariates, gender and country origin (6 countries), were also included into the association tests. 

We focus on gene-based SNP-set tests. Each gene is tested separately; input $p$-values from the group of SNVs within that gene generate a TFisher statistic, then the summary $p$-value of this statistic is obtained to measure how significant the gene is associated. Here we apply the logistic regression model to obtain the input SNV $p$-values, which allows adjusting for other covariates such as non-genetic factors. Specifically, let $y_k$ be the binary indicator of the case ($y_k = 1$) or the control ($y_k=0$) for the $k$th individual, $k = 1, ..., N$. Let $G_k = (G_{k1}, ..., G_{kn})$ denote the genotype vector of $n$ SNVs in the given gene, and let $Z_{k}=(1, Z_{k1}, Z_{k2})$ be the vector of the intercept and covariates of gender and country origin. The logistic regression model is  
\[
{\rm logit}(E(Y_k | G_k, Z_k)) = G_k^{\prime} \beta + Z_k^{\prime} \gamma,
\]
where $\beta$ and $\gamma$ are the coefficients. The null hypothesis is that none of the SNVs in the gene are associated, and thus this gene is not associated: 
\[
H_0: \beta_i = 0, i=1, ..., n.
\]
To test this null hypothesis, we adopt a classic marginal test statistic \citep{mccullagh1989generalized, schaid2002score}
\[
U_i = \sum_{k=1}^N G_{ki}(Y_k-\tilde{Y}_k), \text{ }  i = 1, ..., n,
\]
where $\tilde{Y}_k$ is the fitted probability of the case under $H_0$. It can be shown that under $H_0$, the vector of statistics $U=(U_1, ..., U_n) \overset{D}{\to} N(0, \Sigma)$, as $N \to \infty$, where $\Sigma$ can be estimated by 
\[
\hat{\Sigma} = G^{\prime}WG - G^{\prime}WZ(Z^{\prime}WZ)^{-1}Z^{\prime}WG,
\]
where $G=(G_{ki})$ and $Z=(Z_{ki})$ are the corresponding design matrices, and the diagonal matrix $W=\text{diag}(\tilde{Y}_k(1-\tilde{Y}_k))$. After de-correlation we get the input statistics $X = \hat{\Sigma}^{-\frac{1}{2}}U \overset{D}{\to} N(0, I_{n \times n})$, and the input $p$-values are $2P(N(0,1) > |X_i|) \overset{{\rm i.i.d.}}{\to} \text{Uniform}[0, 1]$. Thus our $p$-value calculation methods given in Section \ref{Sect_P_Calcu} can be applied to any TFisher or oTFisher statistics. 

\begin{figure}[H] \centering%
\begin{tabular}{l}
\includegraphics[width=3in]{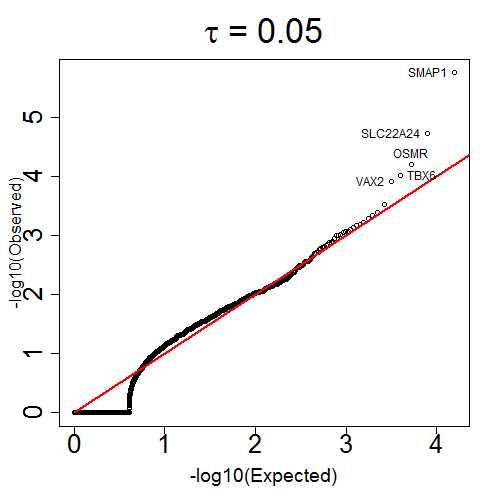}
\includegraphics[width=3in]{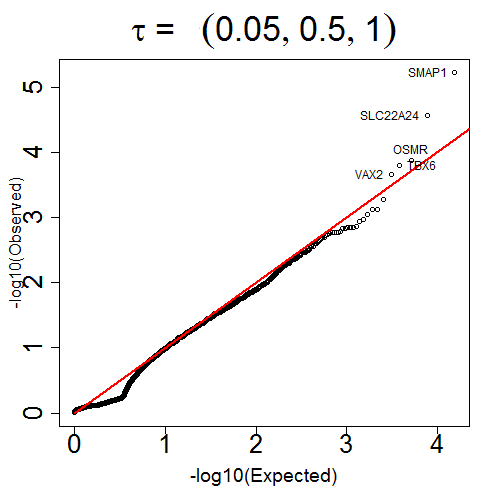}
\end{tabular}%
\caption{Q-Q plots of $p$-values based on soft-thresholding tests. Left: $\tau_1=\tau_2=0.05$. Right: omnibus with $\tau_1=\tau_2 \in \{0.05,0.5,1\}$.}
\label{fig.QQplot}
\end{figure}

The left panel of Figure \ref{fig.QQplot} gives the Q-Q plot of the gene-level $p$-values of TFisher statistics at $\tau_1=\tau_2=0.05$. Because of the truncation, it is natural that some genes have $p$-values at 1 (indicated by the flat part of the dots). It often happens when the gene contains only a few SNVs and their marginal $p$-values, as the input of its TFisher statistic, are all large, say larger than 0.05 here. Such genes are likely not associated anyway, thus the truncation does not influence the type I error rate being well controlled at the gene level. The majority of $p$-values are still along the diagonal as expected. The right panel of Figure \ref{fig.QQplot} provides the Q-Q plot of the gene-level $p$-values by oTFisher test, in which the parameters $\tau_1=\tau_2$  adapt over $ \{0.05,0.5,1\}$. The top ranked genes by both methods are consistent, which is reasonable because the signals, i.e., the ALS associated SNVs, are expected to be in a small proportion of all SNVs. 

To the best of our knowledge, most of these top ranked genes have not been directly reported in genetic association studies of ALS, even though they are promisingly related to ALS from the functionality perspective as discussed below. This result indicates that TFisher tests could likely contribute extra power over existing methods for the discovery of novel disease genes. Certainly, the result is based on very limited data; further statistical and biological validations are needed to clarify their genetic mechanisms to ALS. 

The biological relevance of the top ranked genes  
is briefly discussed here. Gene {\it SMAP1} (a group of 8 SNVs, $p$-value $1.76\times 10^{-6}$) is among significant clusters of altered genes in frontal cortex of ALS samples \citep{andres2017amyotrophic}. The STRING protein-protein network \citep{szklarczyk2014string} shows that it has a strong connection with {\it LRRK2}, a gene associated with late-onset Parkinson's disease (PD), which is a neurodegenerative diseases closely related to ALS \citep{bonifati2006parkinson}. 
Gene {\it SLC22A24} (12 SNVs, $p$-value $1.85\times 10^{-5}$) has reported statistical association with Alzheimer's disease, another neurodegenerative disease closely related to ALS \citep{ayers2016loss}. Furthermore, STRING network shows that {\it SLC22A24} has strong connections with two ALS related genes: {\it AMACR}  and {\it C7orf10}. {\it AMACR} is a gene of AMACR deficiency, a neurological disorder similar as ALS; both initiate and slowly worsen in later adulthood. {\it C7orf10} is associated with ALS types 3 and 4 \citep{fanning2012functional}.
Gene {\it OSMR}    (8 SNVs, $p$-value $6.35\times 10^{-5}$) has been found critically involved in neuronal function regulation and protection \citep{guo2015oncostatin}. Also, it is associated with {\it IL31RA} functional receptor, which is a critical neuroimmune link between TH2 cells and sensory nerves \citep{cevikbas2014sensory}. 
Gene {\it TBX6}     (8 SNVs, $p$-value $9.47\times 10^{-5}$) involves regulation in neural development and maturation \citep{chapman1998three}. Moreover, in a novel stem cell therapy of ALS, {\it TBX6} and its associated {\it SOX2} play a critical role \citep{s2012neuroprotection}. 
Gene {\it VAX2}     (7 SNVs, $p$-value $1.22\times 10^{-4}$) plays a functional role in specifying dorsoventral forebrain. It has direct protein-protein interaction with ALS gene {\it CHMP2B} \citep{cox2010mutations}. It also has a direct STRING connection with {\it SIX3}, which proliferates and differentiates neural progenitor cells (GeneCards database: www.genecards.org). 
Gene {\it GFRA1}    (4 SNVs, $p$-value $2.99 \times 10^{-4}$) encodes a member of the glial cell line-derived neurotrophic factor receptor (GDNFR). It has direct STRING connection with two ALS related genes:  {\it RAP1A}, which is associated with ALS by influencing the activation of {\it Nox2}, a modifier of survival in ALS\citep{carter2009redox}, and {\it PIK3CA}, which is an up-regulated gene in the ALS mouse model \citep{de2014early}.

%%%========================================

\section{Discussion}\label{Sect_Discu}

We proposed and studied a family of Fisher type $p$-value combination tests, TFisher, with a general weighting and truncation scheme, for which many existing methods are special cases.  For the signal detection problem, we studied the optimal TFisher statistics that maximize the BE and the APE. As a result, we showed that soft-thresholding is nearly the best choice, better than the TPM and RTP methods used in a rich literature of applied statistics.  

From the theoretical perspective, the studies of BE and APE revealed the rules for best weighting and truncating input $p$-values in order to best reveal true signals. Our results validated a general principle: when the signals are sparse and strong, more relative weight should be given to the smallest $p$-values; when the signals are dense and weak, a more ``flat" weighting scheme is appropriate. Meanwhile, the original magnitude of $p$-values often need to be downscaled by the parameter $\tau_2 \in (0, 1)$.  We obtained a quantitative relationship between the optimal weighting and truncation scheme and the signal proportion as well as the signal-to-noise ratio. Therefore, this work moved forward the literature, which were mostly based on ad hoc justification and simulation studies. Moreover, this work demonstrated an idea of designing novel powerful statistics by studying the interactive relationship between the statistic-defining parameters and the $H_0 / H_1$-defining parameters. Based on this idea, the statistic family could be further generalized and powerful methods could be obtained for specific testing problems in the future.  

From the practical perspective, the paper provided analytical calculations for both $p$-value and statistical power for a broad family of TFisher statistics under general hypotheses. Data-adaptive omnibus tests could also be applied to real data with unknown signal pattern. A data analysis pipeline for genetic association studies was illustrated, and a list of putative ALS genes were identified and discussed.

\section{Supplementary Materials }

\begin{description}

\item%[Title: Proofs and Supplementary Figures] 

The supplementary materials provide detailed proofs of all lemmas and theorems, and the supplementary figures for statistical power comparisons among data-adaptive methods: oTFisher, ATPM, and ARTP.

\end{description}

\beginsupplement

\subsection{Proofs for Lemmas and Theorems}

\subsubsection{{\bf Proof of Lemma \ref{Thm.C_StaPoint}}}

%\begin{pro}
The first order partial derivative of $c(\theta; \tau_1, \tau_2)$ with respect to $\tau_1$ is
\begin{align*}
\frac{\partial c(\theta; \tau_1, \tau_2)}{\partial\tau_1} \propto 2V_0\frac{\partial \Delta}{\partial\tau_1}  - \Delta\frac{\partial V_0}{\partial\tau_1},
\end{align*}
where 
\begin{align*}
\frac{\partial \Delta}{\partial\tau_1} &= \log\left(\frac{\tau_2}{\tau_1}\right)(D^\prime(\tau_1)-1),\\
\frac{\partial V_0}{\partial\tau_1} &= \left(1-2(1-\tau_1)\left(1+\log\left(\frac{\tau_2}{\tau_1}\right)\right)+(1-2\tau_2)\left(1+\log\left(\frac{\tau_2}{\tau_1}\right)\right)^2\right).
\end{align*}
At $\tau_1=\tau_2$, we have both partials equal to $0$. 

We further examine the partial derivative with respect to $\tau_2$ and then evaluate it at $\tau_2=\tau_1$, 
\begin{align*}
\frac{\partial \Delta}{\partial\tau_2} \Bigr |_{\tau_2=\tau_1} &= \frac{2(D(\tau_1)-\tau_1)}{\tau_1};
\quad{}\frac{\partial V_0}{\partial\tau_2} \Bigr|_{\tau_2=\tau_1} = 2(1-\tau_1); \\
\Delta \Bigr |_{\tau_2=\tau_1} &=\int_{0}^{\tau_1}-\log \left( u\right)(D^\prime(u)-1)du+\log(\tau_1)(D(\tau_1)-\tau_1);\quad{}  V_0 \Bigr|_{\tau_2=\tau_1} = \tau(2-\tau_1).
\end{align*}
Thus 
\begin{align*}
&\frac{\partial c(\theta; \tau_1, \tau_2) }{\partial\tau_2}\Bigr|_{\tau_2=\tau_1}  = 0 \\
\Leftrightarrow &(D(\tau_1)-\tau_1)(2-\tau_1) - (1-\tau_1)\int_0^{\tau_1}-\log (u)(D^\prime(u)-1)du-(D(\tau_1)-\tau_1)(1-\tau_1)\log(\tau_1)=0 \\
\Leftrightarrow &\int_0^{\tau_1}-\log (u)(D^\prime(u)-1)du = (D(\tau_1)-\tau_1)\left(\frac{2-\tau_1}{1-\tau_1}-\log(\tau_1)\right).
\end{align*}
%\end{pro}

\subsubsection{{\bf Proof of Lemma \ref{Lem.mixedModel_noEpsi}}}

%\begin{pro}
The Bahadur efficiency is $c(\theta; \tau_1, \tau_2) = \frac{(E_1-E_0)^2}{V_0}=\frac{\Delta^2}{V_0}$, where $V_0$ is irrelevant to $H_1$, thus to $\epsilon$. On the other hand, $\Delta = \int_{0}^{\tau_1}-\log \left( \frac{u}{\tau_2 }\right)(D^\prime(u)-1)du$. We can show that $\Delta=\epsilon g(\tau_1,\tau_2,\mu)$. This is equivalent to show $D^\prime(u)-1$ has such separability of $\epsilon$.

By (\ref{equ.D}), $D(x)=1-F_1(F_0^{-1}(1-x))$ where $F_0(x)=G_0(x)$ and $F_1(x)=(1-\epsilon) G_0(x) + \epsilon G_1(x;\mu)$. We can further write
\begin{align*}
D(x) &= 1 - (1-\epsilon) G_0(G_0^{-1}(1-x)) - \epsilon G_1(G_0^{-1}(1-x);\mu)\\
& = 1 - (1-\epsilon) (1-x) - \epsilon G_1(G_0^{-1}(1-x);\mu)\\
& = x + \epsilon - \epsilon x - \epsilon G_1(G_0^{-1}(1-x);\mu).\\
D(x) -x & = \epsilon(1 - x - G_1(G_0^{-1}(1-x);\mu)).\\
D^\prime(x) -1 & = \epsilon\left(-1 + \frac{G_1^\prime(G_0^{-1}(1-x);\mu)}{G_0^\prime(G_0^{-1}(1-x))}\right).
\end{align*}
This completes the proof.
%\end{pro}

\subsubsection{{\bf Proof of Theorem \ref{Thm.C_LocalMax}}}

%\begin{pro}
Following Lemma \ref{Thm.C_StaPoint} for the first-order conditions for maximizing $c(\theta; \tau_1, \tau_2)$ in (\ref{equ.c_theta}), note that for the Gaussian mixture model in (\ref{equ.GaussianHypo}), $D^\prime(x)-1 = \epsilon(e^{\mu\Phi^{-1}(1-x)-\mu^2/2}-1)$ and $D(x)-x = \epsilon(1-x-\Phi(\Phi^{-1}(1-x)-\mu))$. Therefore the optimal $\tau_1=\tau_2=\tau^*$ does not depend on $\epsilon$. 

Let $f(\tau) = (D(\tau)-\tau)\left(\frac{2-\tau}{1-\tau}-\log(\tau)\right)+\int_0^{\tau}\log (u)(D^\prime(u)-1)du$. Note that $f(0)=0$, $f^\prime(0)=1-D^\prime(0)>0$. A sufficient condition for the existence of the root $\tau^*$ is \begin{align*}
f(1)=1-D^\prime(1)-\int_0^{\tau}\log (u)(D^\prime(u)-1)du>0\\
\Leftrightarrow \epsilon + \epsilon\int_0^{\tau}\log (u)(e^{\mu\Phi^{-1}(1-x)-\mu^2/2}-1)du<0.
\end{align*} 
This is equivalent to 
\begin{align*}
e^{-\mu^2/2}-\int_0^{1}\log (u)e^{\mu\Phi^{-1}(1-u)}du>0 \Longleftrightarrow \mu>\underline{\mu}=0.84865.
\end{align*}

Next we will examine the the second order derivatives. In a generic form,
\begin{align*}
\frac{\partial^2c(\theta; \tau_1, \tau_2)}{\partial\tau_1\partial\tau_2} =\frac{1}{V_0}\left(\frac{2\Delta^2\frac{\partial V_0}{\partial\tau_1}\frac{\partial V_0}{\partial\tau_2}}{V_0^2} - \frac{2\frac{\partial V_0}{\partial\tau_1}\frac{\partial \Delta}{\partial\tau_2}+\Delta\frac{\partial^2V_0}{\partial\tau_1\partial\tau_2}}{V_0} + 2\Delta\frac{\partial^2\Delta}{\partial\tau_1\partial\tau_2} + 2\frac{\partial \Delta}{\partial\tau_1}\frac{\partial \Delta}{\partial\tau_2}\right).
\end{align*}
Again $\frac{\partial \Delta}{\partial\tau_1} \Bigr |_{\tau_2=\tau_1}=0$ and $\frac{\partial V_0}{\partial\tau_1} \Bigr |_{\tau_2=\tau_1}=0$. We can simplify 

\begin{align*}
\frac{\partial^2c(\theta; \tau_1, \tau_2)}{\partial\tau_1^2}\Bigr |_{\tau_2=\tau_1} &=\frac{1}{V_0}\left( - \frac{\Delta\frac{\partial^2V_0}{\partial\tau_1^2}}{V_0} + 2\Delta\frac{\partial^2\Delta}{\partial\tau_1^2} \right)\Bigr |_{\tau_2=\tau_1},\\
\frac{\partial^2c(\theta; \tau_1, \tau_2)}{\partial\tau_2^2}\Bigr |_{\tau_2=\tau_1} &=\frac{1}{V_0}\left(\frac{2\Delta^2(\frac{\partial V_0}{\partial\tau_2})^2}{V_0^2} - \frac{2\frac{\partial V_0}{\partial\tau_2}\frac{\partial \Delta}{\partial\tau_2}+\Delta\frac{\partial^2V_0}{\partial\tau_2^2}}{V_0} + 2\Delta\frac{\partial^2\Delta}{\partial\tau_2^2} + 2(\frac{\partial \Delta}{\partial\tau_2})^2\right)\Bigr |_{\tau_2=\tau_1},\\
\frac{\partial^2c(\theta; \tau_1, \tau_2)}{\partial\tau_1\partial\tau_2}\Bigr |_{\tau_2=\tau_1} &=\frac{1}{V_0}\left( - \frac{\Delta\frac{\partial^2V_0}{\partial\tau_1\partial\tau_2}}{V_0} + 2\Delta\frac{\partial^2\Delta}{\partial\tau_1\partial\tau_2} \right)\Bigr |_{\tau_2=\tau_1}.
\end{align*}

The following are the relevant terms evaluated at $\tau_2=\tau_1=\tau^*$
\begin{align*}
&\frac{\partial^2V_0}{\partial\tau_1^2} = 2;
\quad{}\frac{\partial^2V_0}{\partial\tau_2^2}=\frac{2(1-\tau^*)}{\tau^*};
\quad{}\frac{\partial^2V_0}{\partial\tau_1\partial\tau_2} = -2;\\
&\frac{\partial^2\Delta}{\partial\tau_1^2} = -\frac{D^\prime(\tau^*)-1}{\tau^*};
\quad{}\frac{\partial^2\Delta}{\partial\tau_2^2}=-\frac{D(\tau^*)-\tau^*}{{\tau^*}^2};
\quad{}\frac{\partial^2\Delta}{\partial\tau_1\partial\tau_2} = \frac{D^\prime(\tau^*)-1}{\tau^*};\\
&\Delta =\frac{(D(\tau^*)-\tau^*)(2-\tau^*)}{1-\tau^*};
\quad{}V_0 = \tau^*(2-\tau^*).
\end{align*}
Plugging them back, we can get the conditions for local maximization. In particular, the condition $\frac{\partial^2c(\theta; \tau_1, \tau_2)}{\partial\tau_1^2}\Bigr |_{\tau_2=\tau_1=\tau*}<0$ is equivalent to $
D(1) < D(\tau^*) + (1-\tau^*)D^\prime(\tau^*)$, which is always true because $D(x)$ is a concave function. Finally, after some algebra, the condition $(\frac{\partial^2c(\theta; \tau_1, \tau_2)}{\partial\tau_1^2}\frac{\partial^2c(\theta; \tau_1, \tau_2)}{\partial\tau_2^2} - (\frac{\partial^2c(\theta; \tau_1, \tau_2)}{\partial\tau_1\partial\tau_2})^2)\Bigr |_{\tau_2=\tau_1=\tau*}>0$ is equivalent to 
 \begin{align*}
\frac{D(\tau^*)-\tau^*}{D^\prime(\tau^*)-1}>2-\tau^*.
\end{align*}
One sufficient condition for such inequality holds is $D^\prime(\tau^*)>1$ which is equivalent to $\tau^*>1-\Phi(\mu/2)$.
%\end{pro}

\subsubsection{{\bf Proof of Theorem \ref{Thm.btheta}}}

%\begin{pro}
Taking the partial of $b(\theta; \tau_1, \tau_2)$ with respect to $\tau_1$, we have
\begin{align*}
\frac{\partial }{\partial \tau_1}b(\theta; \tau_1, \tau_2) & \propto  \left(2V_1\frac{\partial \Delta}{\partial \tau_1} -\Delta\frac{\partial V_1}{\partial \tau_1}\right),
\end{align*}
where 
\begin{align*}
%\frac{\partial V_1}{\partial\tau_1} = &4[\log^2(\tau_1)D^\prime(\tau_1) - 2\log(\tau_1)D^\prime(\tau_1)\int_{0}^{\tau_1}\log (u)D^\prime(u)du \\
\frac{\partial V_1}{\partial\tau_1} = &[\log^2(\tau_1)D^\prime(\tau_1) - 2\log(\tau_1)D^\prime(\tau_1)\int_{0}^{\tau_1}\log (u)D^\prime(u)du \\
+ &2D^\prime(\tau_1)\log(\tau_2)\int_{0}^{\tau_1}\log (u)D^\prime(u)du + 2(D(\tau_1)-1)\log(\tau_2)\log(\tau_1)D^\prime(\tau_1) \\
+ &\log^2(\tau_2)(D^\prime(\tau_1)-2D(\tau_1)D^\prime(\tau_1))].
\end{align*}
Therefore,
\begin{align*}
%\frac{\partial V_1}{\partial \tau_1}|_{\tau_1=\tau_2} = 4[\log^2(\tau_1)D^\prime(\tau_1)-2D^\prime(\tau_1)\log^2(\tau_1)+\log^2(\tau_1)D^\prime(\tau_1)] = 0
\frac{\partial V_1}{\partial \tau_1}\Bigr |_{\tau_2=\tau_1} = [\log^2(\tau_1)D^\prime(\tau_1)-2D^\prime(\tau_1)\log^2(\tau_1)+\log^2(\tau_1)D^\prime(\tau_1)] = 0.
\end{align*}
Together with $\frac{\partial V_0}{\partial \tau_1}\Bigr|_{\tau_1=\tau_2}=\frac{\partial \Delta}{\partial \tau_1}\Bigr|_{\tau_1=\tau_2}=0$, as was shown in the proof of Theorem \ref{Thm.C_LocalMax}, we have
\begin{align*}
\frac{\partial }{\partial \tau_1}b(\theta; \tau_1, \tau_2)\Bigr|_{\tau_1=\tau_2} = 0. 
\end{align*}

The choice $\tau_1=\tau_2=\tau^*$ meets the first order conditions for the optimality if $\frac{\partial }{\partial \tau_2}b(\theta; \tau_1, \tau_2)\Bigr|_{\tau_1=\tau_2=\tau^*} = 0$ has a solution $\tau^*$. This is equivalent to solve
\begin{align*}
\left(2V_1\frac{\partial \Delta}{\partial \tau_2} -\Delta\frac{\partial V_1}{\partial \tau_2}\right)\Bigr|_{\tau_1=\tau_2=\tau^*} = 0, 
\end{align*}
where
\begin{align*}
 \Delta|_{\tau_1=\tau_2} &=-\int_{0}^{\tau_1}\log (u)D^\prime(u)du+D(\tau_1)\log(\tau_1)-\tau_1; 
 \quad{} \frac{\partial \Delta}{\partial \tau_2}|_{\tau_1=\tau_2} = \frac{D(\tau_1)-\tau_1}{\tau_1};\\
%V_1|_{\tau_1=\tau_2} &= 4[ \int_{0}^{\tau_1}\log^2 (u)D^\prime(u)du - (\int_{0}^{\tau_1}\log (u)D^\prime(u)du)^2 \\
V_1|_{\tau_1=\tau_2} &= [ \int_{0}^{\tau_1}\log^2 (u)D^\prime(u)du - (\int_{0}^{\tau_1}\log (u)D^\prime(u)du)^2 \\
&+ 2(D(\tau_1)-1)\log(\tau_1)\int_{0}^{\tau_1}\log (u)D^\prime(u)du +\log^2(\tau_1)D(\tau_1)(1-D(\tau_1))];\\
\frac{\partial V_1}{\partial \tau_2}|_{\tau_1=\tau_2} &= 2\frac{D(\tau_1)-1}{\tau_1}[\int_{0}^{\tau_1}\log (u)D^\prime(u)du-D(\tau_1)\log(\tau_1)].
\end{align*}

Plug them in and after simplification, we want to solve 
\begin{align*}
f_b(\tau) &= (g_1(\tau))^2 - \frac{\tau(1-2\log(\tau))(D(\tau)-1)}{1-\tau}g_1(\tau) - \frac{D(\tau)-\tau}{1-\tau}g_2(\tau) \\
&+ \frac{D(\tau)(D(\tau)-1)\log(\tau)(1-\log(\tau))}{1-\tau} = 0,
\end{align*}
where $g_k(\tau)=g_k(\tau; \epsilon,\mu)=\int_{0}^1\log^k(u)D^\prime(u)du$. 

It is easy to check that $f_b(0)=0$ and $f_b^\prime(0)<0$. A sufficient condition for the existence of a root is that $f_b(1)>0$, i.e.,
\begin{align*}
f_b(1) &= (g_1(1))^2 + D^\prime(1)g_1(1) - (1-D^\prime(1))g_2(1) >0.
\end{align*}

Notice that $g_1(1)=\epsilon\tilde{g}_1(\mu)-1$ and $g_2(1)=\epsilon\tilde{g}_2(\mu)+2$. This is equivalent to 
\begin{align*}
\epsilon[(\tilde{g}_1(\mu))^2-\tilde{g}_1(\mu)-\tilde{g}_2(\mu)]>1+\tilde{g}_1(\mu).
\end{align*}

Since $(\tilde{g}_1(\mu))^2-\tilde{g}_1(\mu)-\tilde{g}_2(\mu)<0$ and $1+\tilde{g}(\mu)$ needs to be $<0$, the sufficient conditions for the existence of a root is 
\begin{align*}
\mu &> \underline{\mu} =0.84865,\\
\epsilon &< \frac{1+\tilde{g}_1(\mu)}{(\tilde{g}_1(\mu))^2-\tilde{g}_1(\mu)-\tilde{g}_2(\mu)},
\end{align*}
where $\underline{\mu}$ is the same given in Theorem~\ref{Thm.C_LocalMax}. 
%\end{pro}

\subsubsection{{\bf Proof of Theorem \ref{Thm.atheta}}}

%\begin{pro}
Taking the partial of $a(\theta; \tau_1, \tau_2)$ with respect to $\tau_1$, we have
\begin{align*}
\frac{\partial }{\partial \tau_1}a(\theta; \tau_1, \tau_2) &\propto (z_\alpha\sqrt{V_0}-\sqrt{n}\Delta)\left( V_1\frac{\partial V_0}{\partial \tau_1} - V_0\frac{\partial V_1}{\partial \tau_1} \right)- \sqrt{n}V_1\left(2V_0\frac{\partial \Delta}{\partial \tau_1} -\Delta\frac{\partial V_0}{\partial \tau_1}\right) \\
& \propto  \left( V_1\frac{\partial V_0}{\partial \tau_1} - V_0\frac{\partial V_1}{\partial \tau_1} \right) -\frac{\sqrt{nV_0}}{z_\alpha}\left(2V_1\frac{\partial \Delta}{\partial \tau_1} -\Delta\frac{\partial V_1}{\partial \tau_1}\right).
\end{align*}

Following the proof of Theorems \ref{Thm.C_LocalMax} and \ref{Thm.btheta}, we have $\frac{\partial V_1}{\partial \tau_1}|_{\tau_1=\tau_2} = \frac{\partial V_0}{\partial \tau_1}|_{\tau_1=\tau_2}=\frac{\partial \Delta}{\partial \tau_1}|_{\tau_1=\tau_2}=0$, and thus
\begin{align*}
\frac{\partial }{\partial \tau_1}a(\theta; \tau_1, \tau_2)\Bigr|_{\tau_1=\tau_2} = 0. 
\end{align*}

The choice $\tau_1=\tau_2=\tau^*$ meets the first order conditions for the optimality if $\frac{\partial }{\partial \tau_2}a(\theta; \tau_1, \tau_2)|_{\tau_1=\tau_2} = 0$ has a solution $\tau^*$. This is equivalent to solve
\begin{align*}
\left[\left( V_1\frac{\partial V_0}{\partial \tau_2} - V_0\frac{\partial V_1}{\partial \tau_2} \right) -\frac{\sqrt{nV_0}}{z_\alpha}\left(2V_1\frac{\partial \Delta}{\partial \tau_2} -\Delta\frac{\partial V_1}{\partial \tau_2}\right)\right]\Bigr|_{\tau_1=\tau_2} = 0, 
\end{align*}
where $V_0|_{\tau_1=\tau_2} = \tau_1(2-\tau_1)$, $\frac{\partial V_0}{\partial \tau_2}|_{\tau_1=\tau_2} = 2(1-\tau_1)$, and the rest of the terms are given in the proof of Theorem~\ref{Thm.btheta}. We can simplify the equation to be
\begin{align*}
f_c(\tau) &= (\tau-c_\tau)(g_1(\tau))^2 - \frac{\tau(D(\tau)-1)}{1-\tau}\left(2\log(\tau)(1-\tau+c_\tau)-2+\tau-c_\tau\right)g_1(\tau)\\
&+  \left(\tau-c_\tau\frac{D(\tau)-\tau}{1-\tau}\right)g_2(\tau) + \frac{\tau D(\tau)(D(\tau)-1)\log(\tau)}{1-\tau}\left(\log(\tau)(1-\tau+c_\tau)-2+\tau-c_\tau\right) = 0,
\end{align*}
where $c_\tau=c_n\sqrt{\tau(2-\tau)}$. Here we have $f_c(0)=0$ and $f_c^\prime(0)>0$. The condition $f_c(1)<0$ means
\begin{align*}
&(1-c_n)\epsilon\left((\tilde{g}_1(1))^2-\tilde{g}_1(1)-\tilde{g}_2(1)\right) - (1-c_n)(\tilde{g}_1(1)+1) \\
+& \epsilon\left(2\tilde{g}_1(1)+\tilde{g}_2(1)\right) - \left(2\tilde{g}_1(1)+\tilde{g}_2(1)\right) < 0.
\end{align*}
For $c_n$ large enough, $(1-c_n)[(\tilde{g}_1(\mu))^2-\tilde{g}_1(\mu)-\tilde{g}_2(\mu)]+2\tilde{g}_1(\mu)+\tilde{g}_2(\mu)>0$.Thus, a sufficient conditions for the existence of $\tau^*$, i.e., the stationary point is 
\begin{align*}
\mu &> \underline{\mu}^\prime\text{ such that } (1-c_n)[1+\tilde{g}_1(\underline{\mu}^\prime)]+2\tilde{g}_1(\underline{\mu}^\prime)+\tilde{g}_2(\underline{\mu}^\prime) = 0, \\
\epsilon &< \frac{(1-c_n)[1+\tilde{g}_1(\mu)]+2\tilde{g}_1(\mu)+\tilde{g}_2(\mu)}{(1-c_n)[(\tilde{g}_1(\mu))^2-\tilde{g}_1(\mu)-\tilde{g}_2(\mu)]+2\tilde{g}_1(\mu)+\tilde{g}_2(\mu)}.
\end{align*}

%\end{pro}

\subsection{Supplementary Figures}

\begin{figure}[H] \centering%
\begin{tabular}{l}
\includegraphics[width=2in]{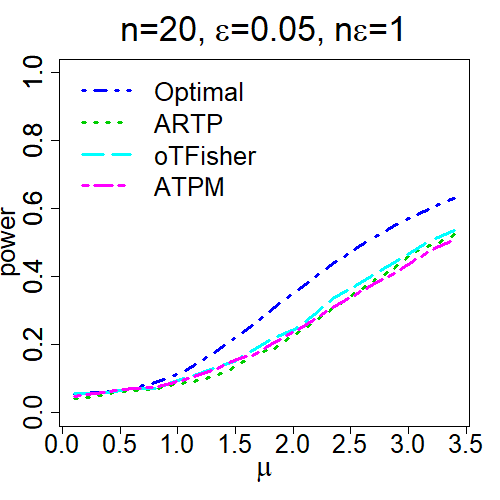}
\includegraphics[width=2in]{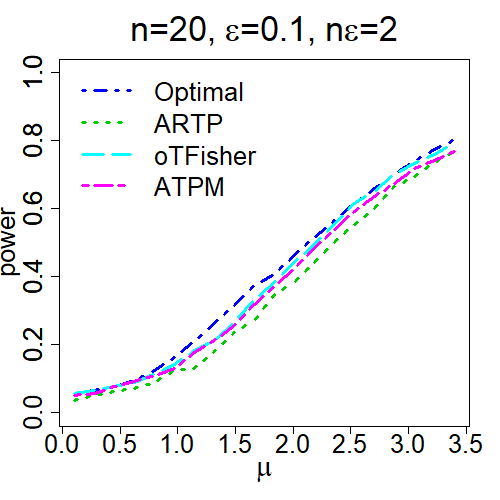}
\includegraphics[width=2in]{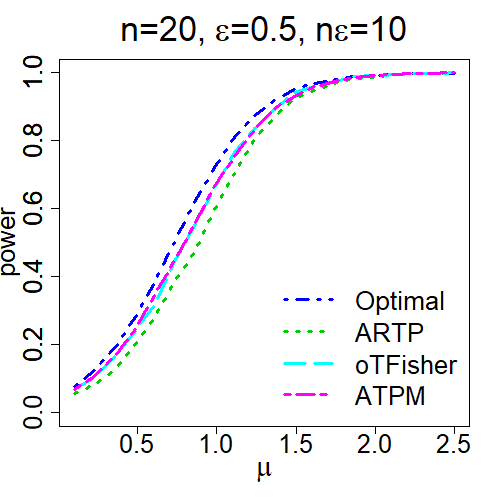}\\
\includegraphics[width=2in]{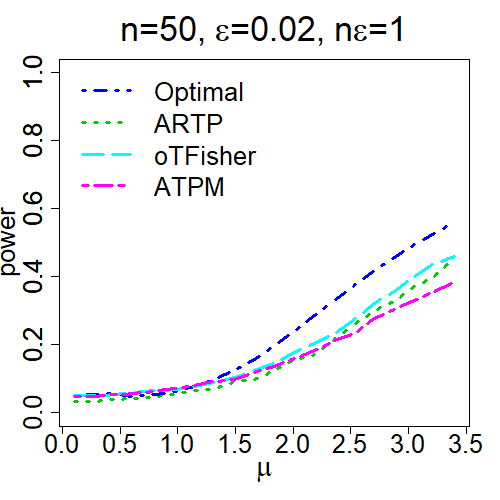}
\includegraphics[width=2in]{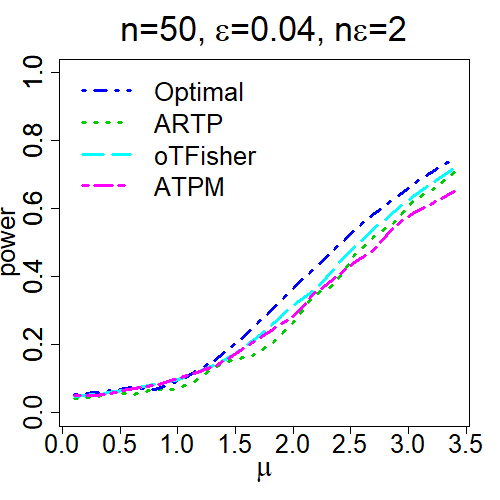}
\includegraphics[width=2in]{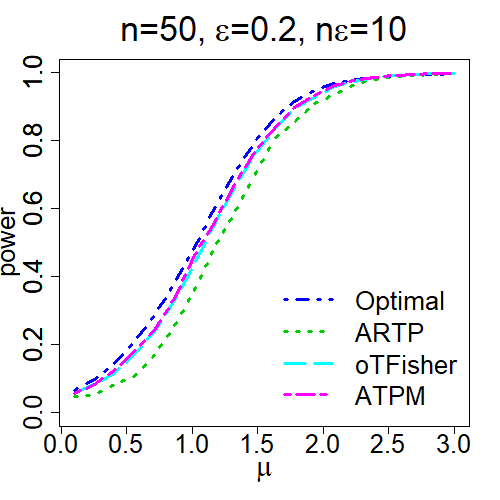}\\
\includegraphics[width=2in]{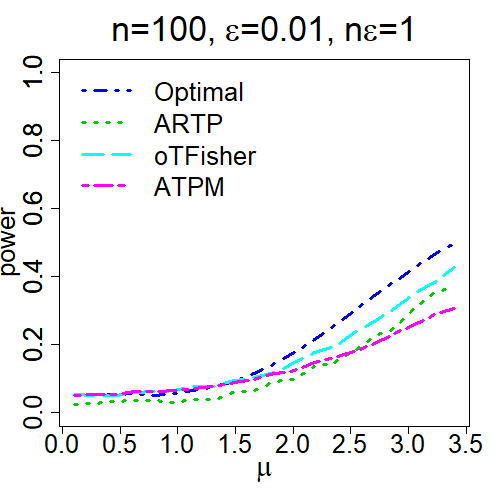}
\includegraphics[width=2in]{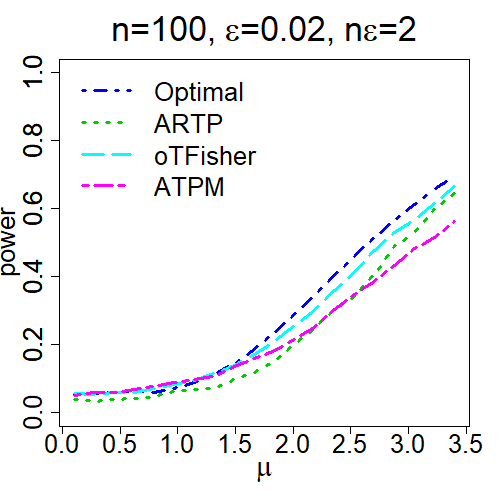}
\includegraphics[width=2in]{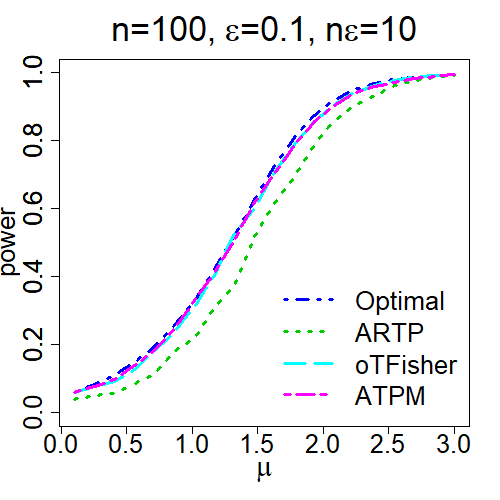}
\end{tabular}%
\caption{Power comparison between the optimal and adaptive tests over signal strength $\mu$. Type I error rate $\alpha=0.05$. Optimal: optimal TFisher at maximizers $\tau_1^*,\tau_2^*$ of APE; ARTP: adaptive RTP with adaptive $K\in\{1, 0.05n,0.5n,n\}$; oTFisher: soft-thresholding omnibus TFisher with adaptive $\tau \in \{0.01,0.05,0.5, 1\}$; ATPM: adaptive TPM (hard-thresholding) with adaptive $\tau \in \{0.01,0.05,0.5, 1\}$.}
\label{fig.artppower_over_mu}
\end{figure}

\begin{figure}[H] \centering%
\begin{tabular}{l}
\includegraphics[width=2in]{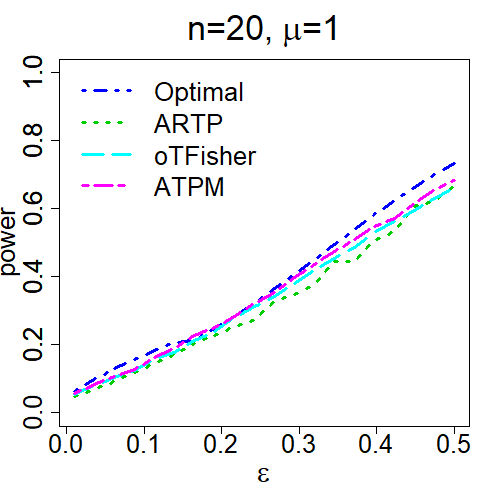}
\includegraphics[width=2in]{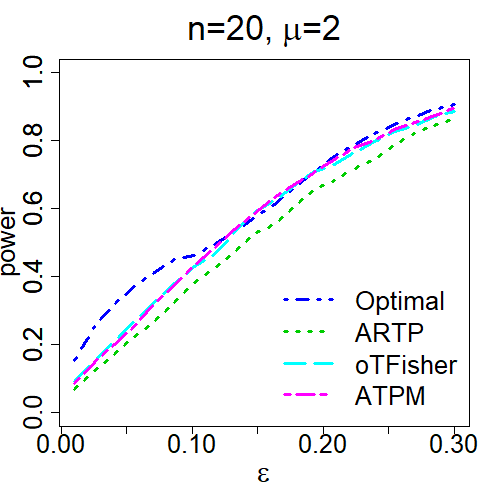}
\includegraphics[width=2in]{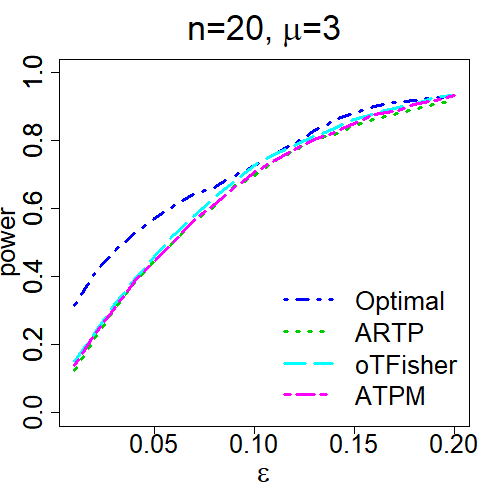}\\
\includegraphics[width=2in]{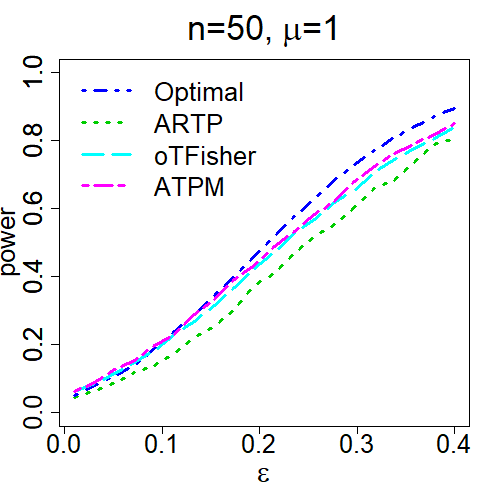}
\includegraphics[width=2in]{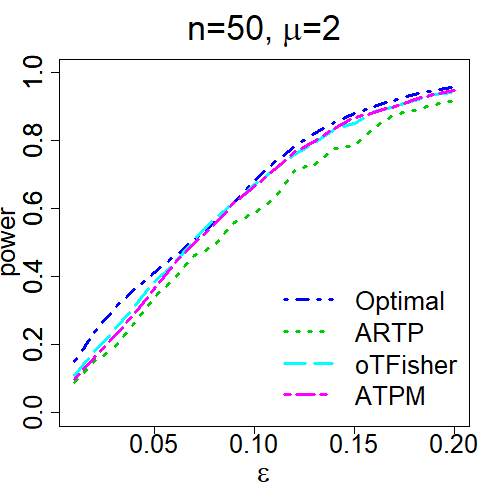}
\includegraphics[width=2in]{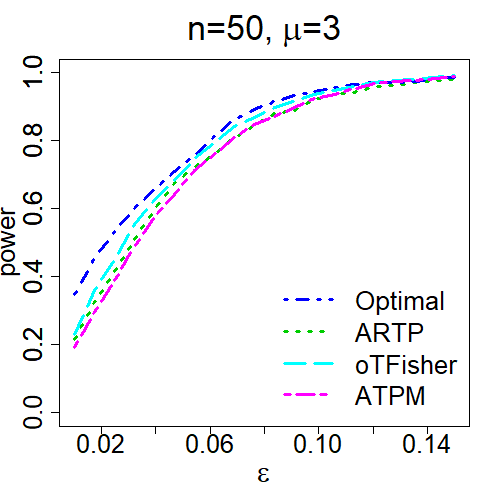}\\
\includegraphics[width=2in]{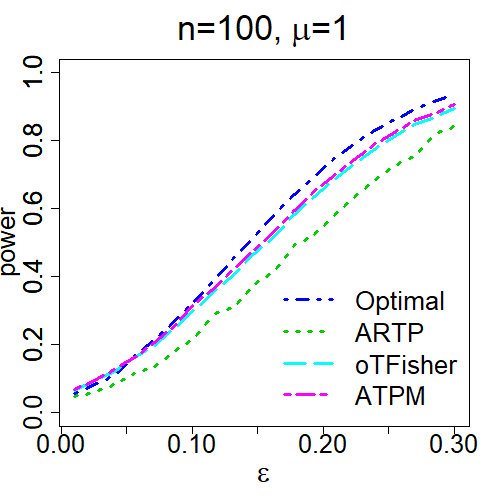}
\includegraphics[width=2in]{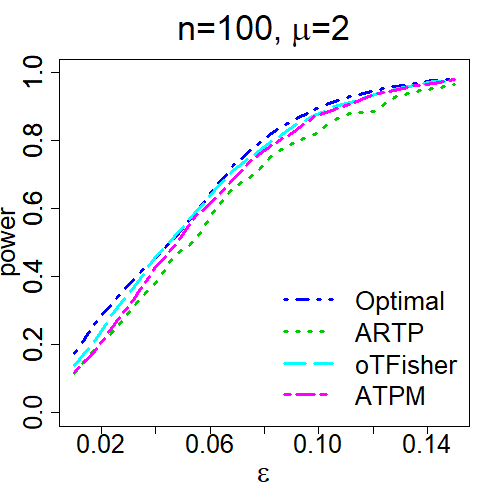}
\includegraphics[width=2in]{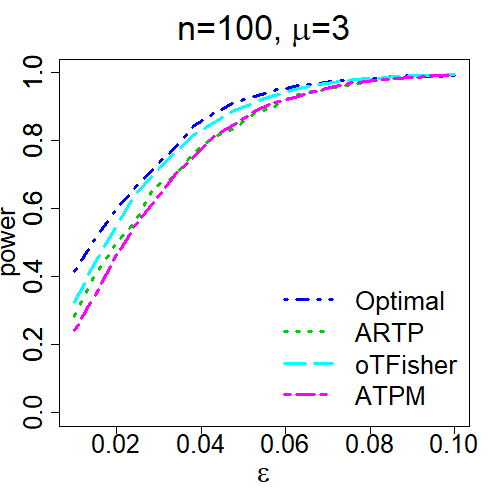}
\end{tabular}%
\caption{Power comparison between the optimal and adaptive tests over signal proportion $\epsilon$. Type I error rate $\alpha=0.05$. Optimal: optimal TFisher at maximizers $\tau_1^*,\tau_2^*$ of APE; ARTP: adaptive RTP with adaptive $K\in\{1, 0.05n,0.5n,n\}$; oTFisher: soft-thresholding omnibus TFisher with adaptive $\tau \in \{0.01,0.05,0.5, 1\}$; ATPM: adaptive TPM (hard-thresholding) with adaptive $\tau \in \{0.01,0.05,0.5, 1\}$.}
\label{fig.artppower_over_eps}
\end{figure}

%% HERE WE DECLARE THE BIBLIOGRAPHYSTYLE TO USE AND THE BIBLIOGRAPHY DATABASE
\bibliography{allMyReferences}
%\bibliography{example}
%\bibliographystyle{ECA_jasa}
%\bibliographystyle{ieeetr}

\end{document}